\documentclass[aps,prd,twocolumn,showpacs,superscriptaddress,groupedaddress]{revtex4}
\usepackage{amsmath}
\usepackage{latexsym}
\usepackage{amsfonts}
\usepackage{graphicx}
\usepackage{subfigure}
\usepackage{mathrsfs}
\usepackage{epstopdf}
\usepackage{ulem}
\usepackage{bm}
\usepackage{color}
\usepackage[colorlinks,linkcolor=blue,citecolor=blue]{hyperref}

\begin{document}

\title{Constraints on the dark matter and dark energy interactions from weak lensing bispectrum tomography}

\author{Rui An}
\email{an\_rui@sjtu.edu.cn}
\affiliation{School of Physics and Astronomy, Shanghai Jiao Tong University,
Shanghai 200240, China}

\author{Chang Feng}
\email{chang.feng@uci.edu}
\affiliation{Department of Physics and Astronomy,
University of California, Irvine, CA 92697, USA }

\author{Bin Wang}
\email{wang\_b@sjtu.edu.cn}
\affiliation{School of Physics and Astronomy, Shanghai Jiao Tong University,
Shanghai 200240, China}

\begin{abstract}

We estimate uncertainties of cosmological parameters for phenomenological interacting dark energy models using weak lensing convergence power spectrum and bispectrum. We focus on the bispectrum tomography and examine how well the weak lensing bispectrum with tomography can constrain the interactions between dark sectors, as well as other cosmological parameters. Employing the Fisher matrix analysis, we forecast parameter uncertainties derived from weak lensing bispectra with a two-bin tomography and place upper bounds on strength of the interactions between the dark sectors. The cosmic shear will be measured from upcoming weak lensing surveys with high sensitivity, thus it enables us to use the higher order correlation functions of weak lensing to constrain the interaction between  dark sectors and will potentially provide more stringent results with other observations combined.

\end{abstract}

\maketitle

\section{Introduction}

It is widely believed that our Universe is undergoing an accelerated expansion. This prevailing understanding was obtained from independent observations, including the supernovae type Ia (SNIa) \cite{Perlmutter1998,Perlmutter1999,Riess1998,Riess1999}, temperature anisotropies of the cosmic microwave background (CMB) \cite{Spergel2003, Planck2014I,Planck2014XX}, inhomogeneities in the matter distribution \cite{Tegmark2004,Cole2005}, the integrated Sachs-Wolfe (ISW) effect \cite{Boughn2004}, baryon acoustic oscillations (BAO) \cite{Eisenstein2005}, weak lensing \cite{Contaldi2003} and gamma-ray bursts \cite{Kodama2008}. Within the framework of Einstein gravity, the acceleration is driven by dark energy which is a new energy component with negative pressure. The standard $\Lambda$CDM model is considered as the most favorable model in driving the present acceleration and it is proven to be consistent with a lot of observational evidence, although recently the $\Lambda$CDM model was found to have trouble in explaining the mismatch between the Hubble constant $H_0$ directly inferred from a local measurement \cite{Riess2016} and the CMB experiment \cite{Planck2016}. However, the $\Lambda$CDM model is not satisfactory from the theoretical point of view. It suffers a cosmological constant problem, i.e., the observed value is many orders of magnitude smaller than the prediction of quantum field theory \cite{Weinberg1989}. In addition, a coincidence problem \cite{Zlatev1999} is also associated with the $\Lambda$CDM model in which the dark energy is construed as a constant that is difficult to explain why the dark energy plays an important role and why the Universe is accelerating just now but neither earlier nor later. Because of these problems, there are a lot of attempts to find a more preferable model which can substitute the standard $\Lambda$CDM model and explain the late time accelerated expansion of our Universe.

Today we know that the Universe is composed of nearly $25\%$ of dark matter and $70\%$ of dark energy. As enlightened by the field theory, it is natural to consider whether these two major components in the Universe could have some mutual interactions rather than evolve separately. Since neither do we know physics of dark matter nor dark energy to date, interactions between dark sectors can only be explored phenomenologically by introducing energy flow or momentum flow between them. It was argued that some appropriate interactions can alleviate the coincidence problem and understand why energy densities of dark sectors are at the same order of magnitude as observed today \cite{Amendola2000,Pavon2005,Boeher2008,Olivares2006,Chen2008}. For a recent review on the interaction between dark matter and dark energy, please refer to \cite{Wang2016} and references therein.

In addition to theoretical motivations, a viable model of interaction between dark matter and dark energy should pass examinations from observations. In the recent review \cite{Wang2016}, some phenomenological interaction models have been found to reproduce consistent results with the observational data \cite{Feng2007I} in terms of the expansion history of the Universe, the CMB temperature anisotropies including the ISW effect \cite{Olivares2008,Schaefer2008,Xia2009,Matsubara2004}, the kinetic Sunyaev-Zeldovich (kSZ) effect \cite{Xu2013} , BAO, SNIa, Hubble constant \cite{Amendola2007,Feng2007II,Bean2008,Micheletti2009}, and galaxy clustering \cite{Abdalla2009,Abdalla2010}. As complimentary tests, it is of great interest to confront the interaction models between dark sectors to new probes. The interaction between dark sectors can change time evolution of gravitational potential which results in modification of low-multipoles of the radiation power spectrum. Also, the interaction can change the gravitational potential and modulate peculiar velocities of baryons which in turn leave footprint on temperature fluctuations of the CMB via the kSZ effect at small scales. Moreover, the change of the gravitational potential induced by the interaction can also deflect trajectories of photons emitted by distant objects which would appear to be slightly distorted, giving birth to a weak gravitational lensing effect. The interaction between dark sectors can alter all the information encoded in the distance-redshift relation, the matter distribution in the Universe, as well as the growth of density perturbations observed from lensed objects. Weak lensing can not only become important in probing dark energy  \cite{Huterer2002}\cite{Takada2004}, but also serve as a complementary probe to determining the interaction between dark sectors.

The dependence of weak lensing power spectrum on the interaction between dark sectors was first discussed in \cite{Vacca2008}. However, the specific mechanism of the interacting dark fluids was not clearly investigated in their study. Bispectra are advantageous in that it is much easier to carry out a perturbative expansion and they are sensitive to transitions from linear to nonlinear dynamics in structure formation. In light of this, a supplemental study focused on the bispectrum tomography using an interesting model that allows dark matter to decay into dark energy \cite{Schafer2008}. Some predictions were also presented for non-linear weak lensing from numerical simulations, assuming a coupling between a dark energy scalar field and cold dark matter fluid \cite{Beynon2012, Giocoli2015}.

For more general phenomenological interacting dark energy models, the effects of different interactions between dark sectors can alter redshift evolutions of mass clustering and distance-redshift relation in different ways. It is interesting to investigate how lensing tomography with bispectrum and power spectrum can constrain cosmological parameters and how the constraints of weak lensing might differ from other observations. We will follow the phenomenological principles and will not restrict ourselves to specific models of dark energy, but consider a constant or time-dependent equation of state (EoS) $w$ for dark energy in multiple regions, e.g., $w<-1$ or $-1<w<-1/3$. We will pay much attention to the weak lensing bispectrum, because future wide field surveys promise to measure it at high significance and it can provide additional information on structure formation that can not be extracted from the power spectrum. We will study all the parameters to which the lensing power spectrum and bispectrum are sensitive. Especially, we will constrain the strength of the interaction in different interaction models between dark sectors and examine its degeneracy with other parameters.

The structure of the paper is as follows: we briefly introduce phenomenological models of dark matter and dark energy in Section II where we incorporate the interactions between  dark sectors. In Section III, we derive power spectrum, bispectrum and cross-spectra of the weak lensing convergence. In Section IV, we discuss the Fisher matrix formalism for weak lensing bispectrum tomography, and show various constraints on parameters introduced by our models. We discuss the results and conclude in the last section.

\section{Dark matter and dark energy interaction}

We will concentrate on the phenomenological model of the interaction between dark matter and dark energy, which is in a linear combination of energy densities of dark sectors \cite{Wang2016}. In these models, the total energy density of dark sectors is conserved, while the energy densities of dark matter and dark energy individually can evolve as
\begin{gather}
\label{eq.rhodm}
\dot{\rho_{c}}+3H\rho_{c}=Q,\\
\label{eq.rhode}
\dot{\rho_{d}}+3H(1+w)\rho_{d}=-Q,
\end{gather}
where $H$ is the Hubble function defined as $H=\dot{a}/a$, $a$ is the scale factor and the dot is the derivative with respect to the conformal time, $w$ is the equation of state for dark energy, and $Q$ represents the interaction kernel.  The sign of $Q$ determines the direction of the energy flow,  the positive sign indicating the energy flow from dark energy to dark matter while the negative sign signaling the opposite. The exact form of $Q$ can not be derived from first principles since we are lack of fundamental theory. The interaction introduces only a small correction to the evolution history of the Universe. In analogy to particle physics, one would expect $Q$ to be a function of both the energy densities $\rho_c, \rho_d$ and the inverse of the Hubble constant $H^{-1}$. To the first order of Taylor expansion, we have $Q=3\lambda_{1}H\rho_{c}+3\lambda_{2}H\rho_{d}$, where $\lambda_1$ and $\lambda_2$ are free parameters to be determined from observations. In Table.\ref{tab.model} we present four phenomenological interacting models \cite{Wang2016} that will be considered in this work to better understand the effect of the interaction. We study the constant equation of state of dark energy in the phantom and quintessence regions respectively to ensure stable density perturbations \cite{Wang2016}. Considering that the time dependent dark energy equation of state can secure stability of the linear perturbations, we will generalize our investigation to equation of state of the form  $w(a)=w_{0}+w_{a}(1-a)$ where $w_0$ and $w_a$ are also free parameters.

\begin{table}[ht]
\caption{Phenomenological interacting dark energy models \label{tab.model}}
\begin{tabular}{ccc}
\hline
Model & $Q$ & $w$\\
\hline
I & $3\lambda_{2}H\rho_{d}$ & $-1<w<-1/3$ \\
II & $3\lambda_{2}H\rho_{d}$ & $w<-1$ \\
III & $3\lambda_{1}H\rho_{c}$ & $w<-1$ \\
IV & $3\lambda H(\rho_{c}+\rho_{d})$ & $w<-1$ \\
V & 0 & $w<-1/3$ \\
\hline
\end{tabular}
\end{table}

In a linear theory, matter density perturbation is defined as $\delta_c=(\rho_c-\langle\rho_c\rangle)/\langle\rho_c\rangle$ and can be derived from its present value via $\delta_c(\bm{k},a)=D(a)\delta_c(\bm{k},1)$, where $D(a)$ is the growth factor with the present value normalized to unity. And the growth function $D(a)$ is solved from the differential equation
\begin{equation}
\label{eq.D}
\begin{split}
&\frac{\text{d}^2D}{\text{d}a^2}+\frac{1}{a}\left[\frac{3}{2}-\frac{3}{2}w(1-\Omega_c)+3\lambda_1+6\frac{\lambda_2}{r}\right]\frac{\text{d}D}{\text{d}a}\\
&=\frac{1}{a^2}\left[\frac{3}{2}\Omega_{c}-3\frac{\lambda_2}{r}\left(2+3\lambda_1+3\frac{\lambda_2}{r}-\frac{\ln r}{\ln a}+\frac{\text{d}\ln H}{\text{d}\ln a}\right)\right]D,
\end{split}
\end{equation}
where $r=\rho_c/\rho_d$, and $\Omega_c=\rho_c/\rho_{\rm{crit}}$ with the critical density defined as $\rho_{\rm{crit}}=3H^2/(8\pi G)$. Here we do not consider the dark energy perturbation in  Eq. (\ref{eq.D}) since its influence is very small compared to dark matter perturbation \cite{He2009}. This is a general expression in describing  the growth of structure for the interacting dark energy models.  If we neglect the interaction between dark sectors by setting $\lambda_1=\lambda_2=0$ and keep $w$ time dependent, Eq. (\ref{eq.D}) will reduce to Eq. (4) in \cite{Takada2004} for evolving dark energy models. When we set $w=-1$ and $\lambda_1=\lambda_2=0$, Eq. (\ref{eq.D}) will reduce to Eq. (6) in \cite{Schafer2008}, which is valid only for the standard $\Lambda$CDM model. In \cite{Schafer2008}, the evolution of  the standard growth function was assumed to be valid even in the decaying cold dark matter model as the decay was not expected to affect the overdensity field significantly.

\begin{figure*}
\includegraphics[scale=0.3]{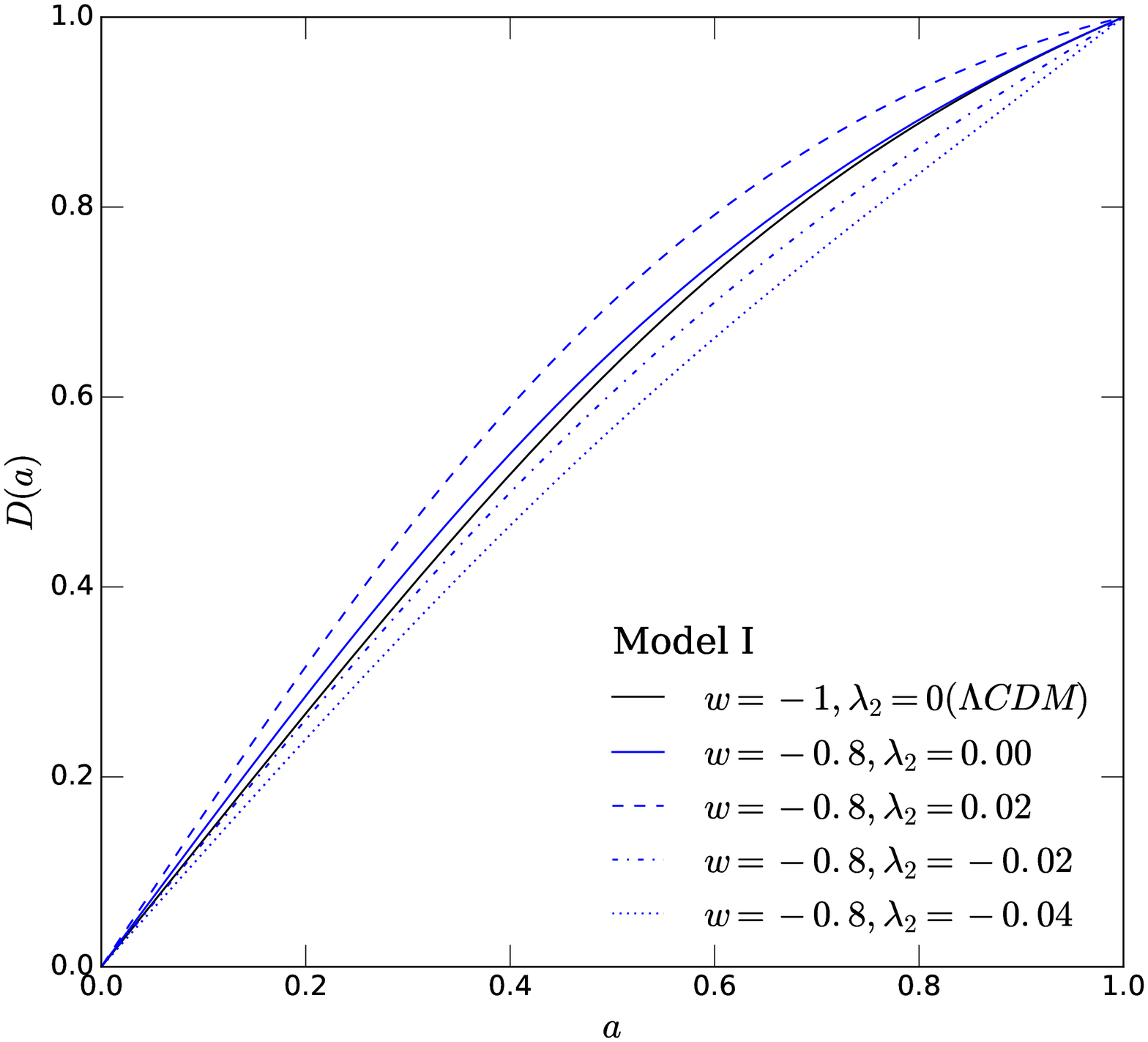}
\includegraphics[scale=0.3]{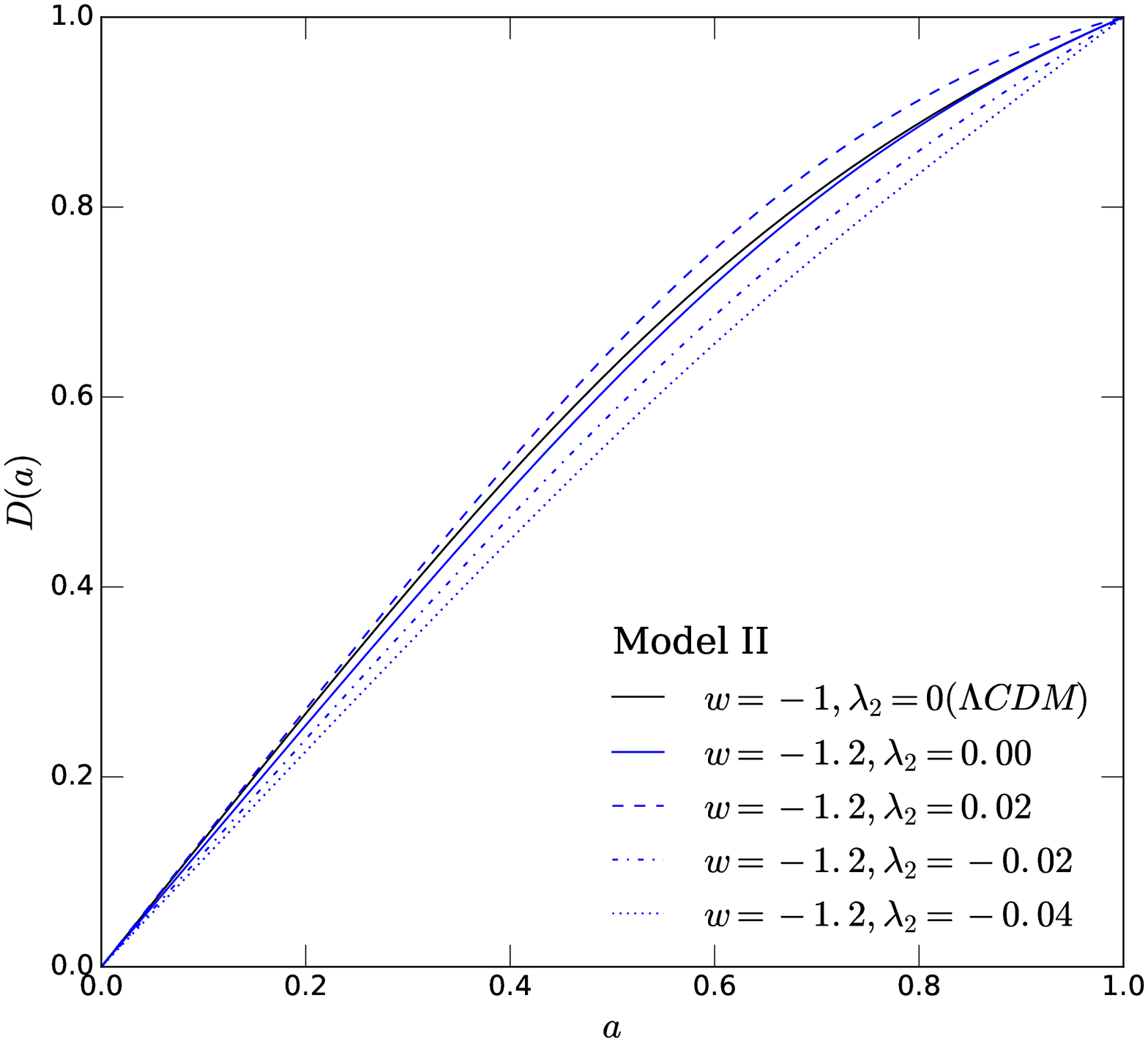}
\includegraphics[scale=0.3]{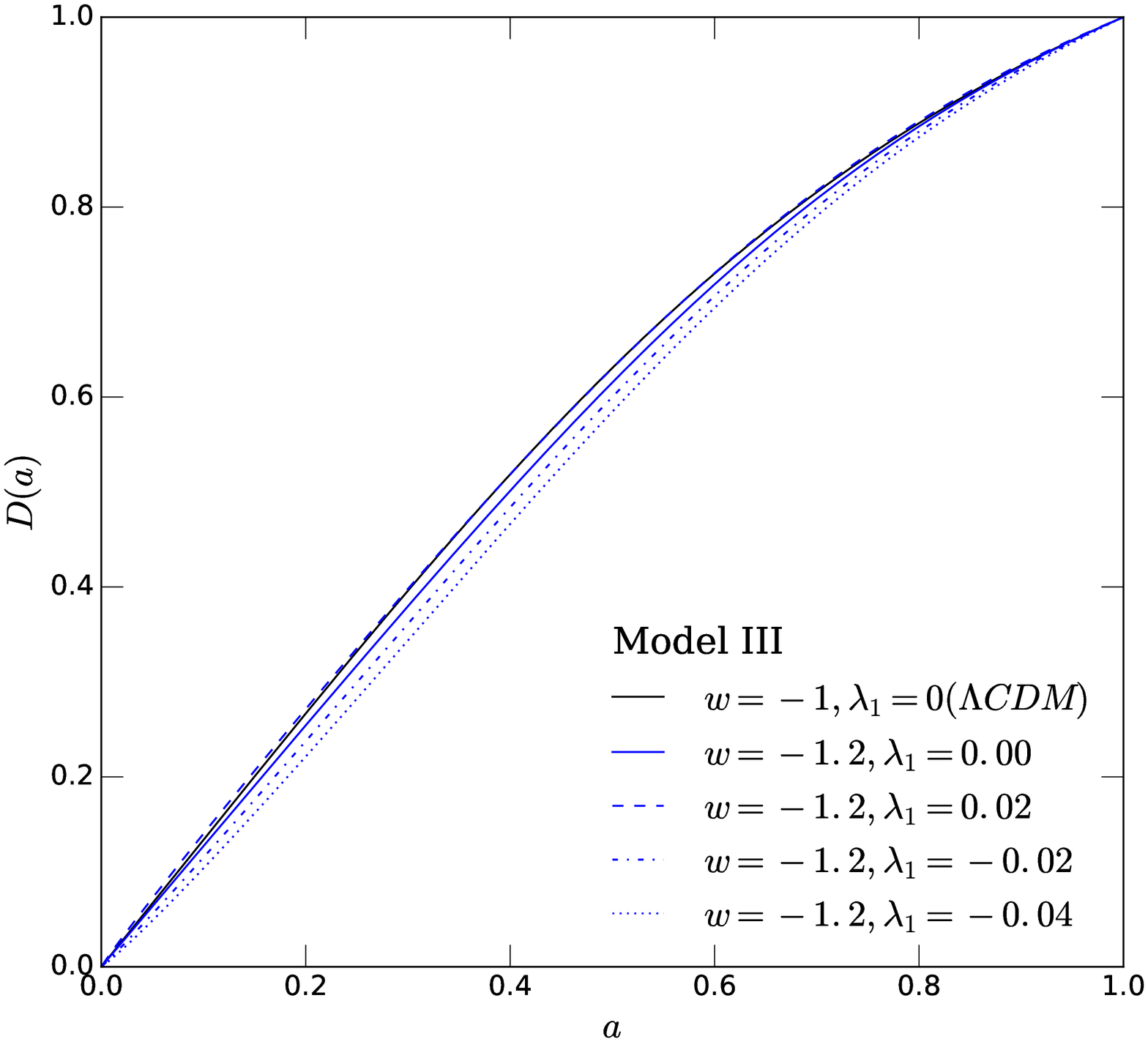}
\includegraphics[scale=0.3]{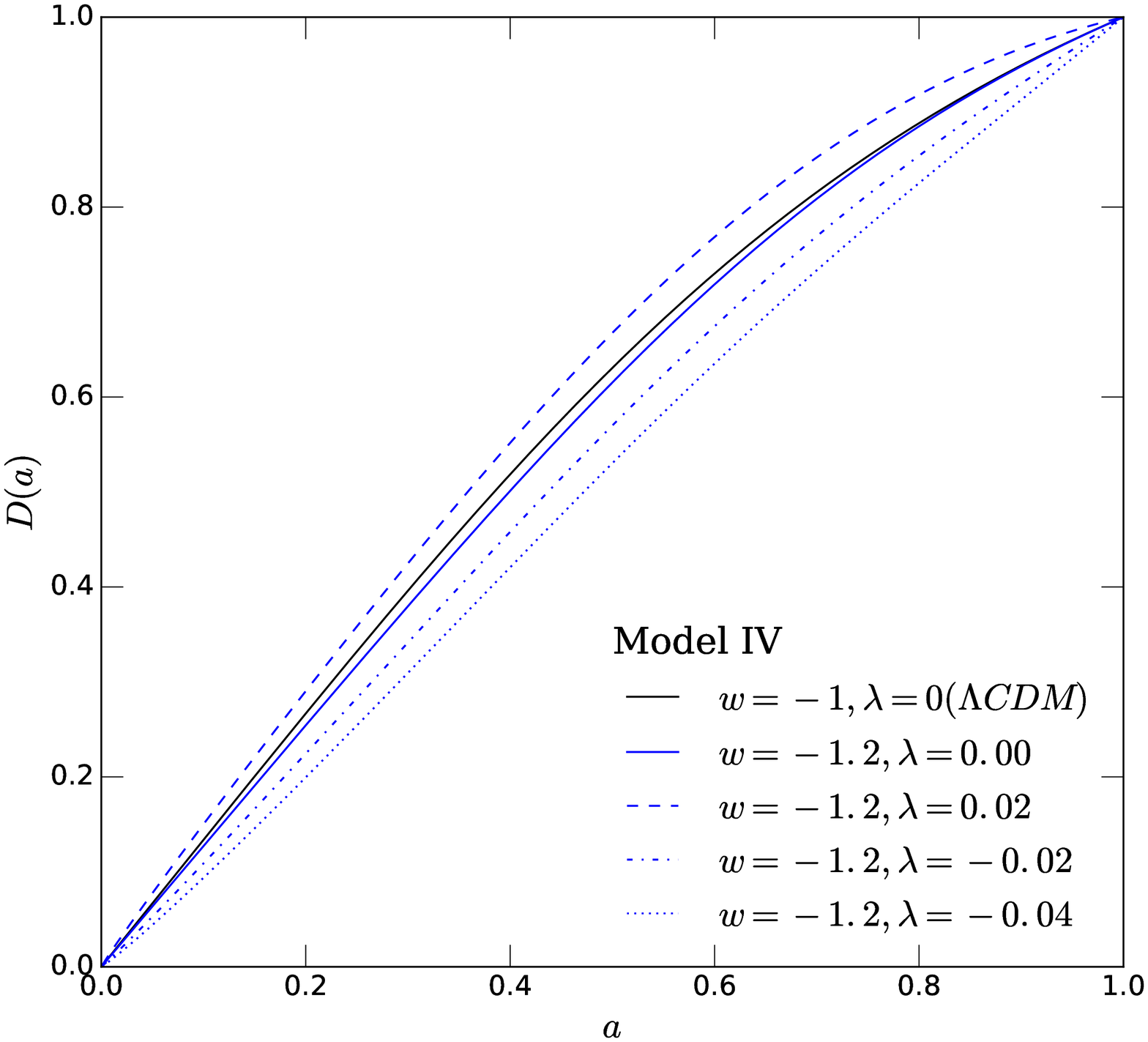}
\caption{Growth function D(a) for interacting dark energy models with constant equation of state.}
\label{fig.D}
\end{figure*}

Solving the general differential equation in Eq. (\ref{eq.D}),  we illustrate evolutions of the growth function for the models listed in Table \ref{tab.model} with constant equation of state in Fig. (\ref{fig.D}). In the non-interacting dark energy models (Model V), the standard $\Lambda$CDM model has lower redshift evolution of the growth factor of the Universe than $w>-1$ cases, but a higher value of the matter density growth than $w<-1$ cases \cite{Takada2004,Linder2003}. For all the models with $Q < 0$, the dark matter clustering source term $\Omega_c$ was large in the past and it was able to develop the same structure although the growth factor was lower. For all the models with $Q > 0$, they require larger growth factors to compensate the lack of dark matter in the past and form the same structure today. For models II, III and IV, the degeneracy between the interaction parameters and $w$ can further complicate the evolution of the growth factor. With $w < -1$ in models II, III and IV, the redshift evolution of the growth rate was faster and the growth factor was suppressed due to the effect of both dissipation and source terms if there is no interaction between dark sectors. The growth factor could be either enhanced or suppressed when $Q > 0$ or $Q < 0$, respectively. The coupling effect is more prominent in models I and IV.

For the simple parameterization of the time evolving equation of state, we find that when $w > -1$, the growth factor becomes higher in the redshift space, while $w < -1$ has the opposite effect. For conciseness, we do not illustrate the result for the time dependent equation of state of dark energy.

\section{Weak lensing}

Interactions between dark sectors can change both the growth rate and the density of the cold dark matter. Eventually the weak lensing convergence will become dependent of this tiny interaction. The convergence field $\kappa(\bm{\theta})$ is given by the projected mass density along the line of sight \cite{Bartelmann2001,Mellier1999,Hoekstra2008}
\begin{equation}
\label{eq.kappa}
\kappa(\bm{\theta})=\int_{0}^{\chi_H}d\chi W(\chi)\delta[\chi \bm{\theta},\chi],
\end{equation}
where $\bm{\theta}$ is a angular position on the sky and $\chi_H$ is a comoving distance at an infinitely large redshift. The weak lensing weighting function $W(\chi)$ is \cite{Schafer2008}
\begin{equation}
\label{eq.W1}
W(\chi)=\frac{3}{2c^2}a(\chi)^2H(\chi)^2\Omega_c(\chi)\chi \int_{\chi}^{\chi_H}d\chi'p(z)\frac{dz}{d\chi'}\frac{\chi'-\chi}{\chi'},
\end{equation}
where $p(z)$ is a normalized redshift distribution of lensing galaxies and $c$ is the speed of light. Here we use the distribution \cite{Bartelmann2001}
\begin{equation}
\label{eq.p}
p(z)=\frac{\beta}{\Gamma(\frac{3}{\beta})}\Big(\frac{z^2}{z_0^3}\Big)\exp[-\Big(\frac{z}{z_0}\Big)^{\beta}].
\end{equation}
This expression is normalized and provides a good fit to the observed redshift distribution \cite{Smail1995}. The parameters are $\beta=3/2$ and $z_0=0.64$ corresponding to  a median redshift $z_{\rm{med}}=0.9$.

\begin{figure*}
\includegraphics[scale=0.3]{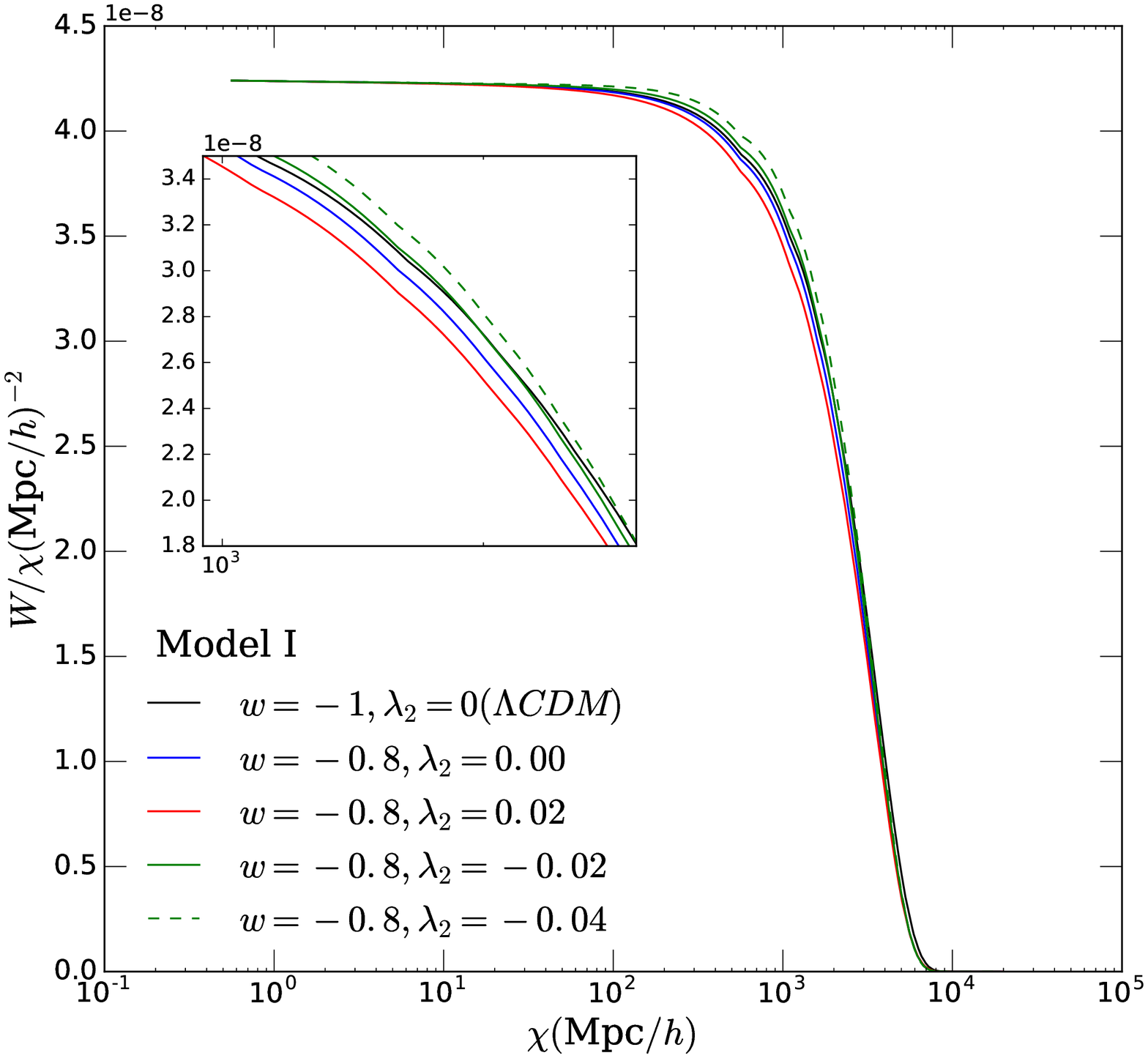}
\includegraphics[scale=0.3]{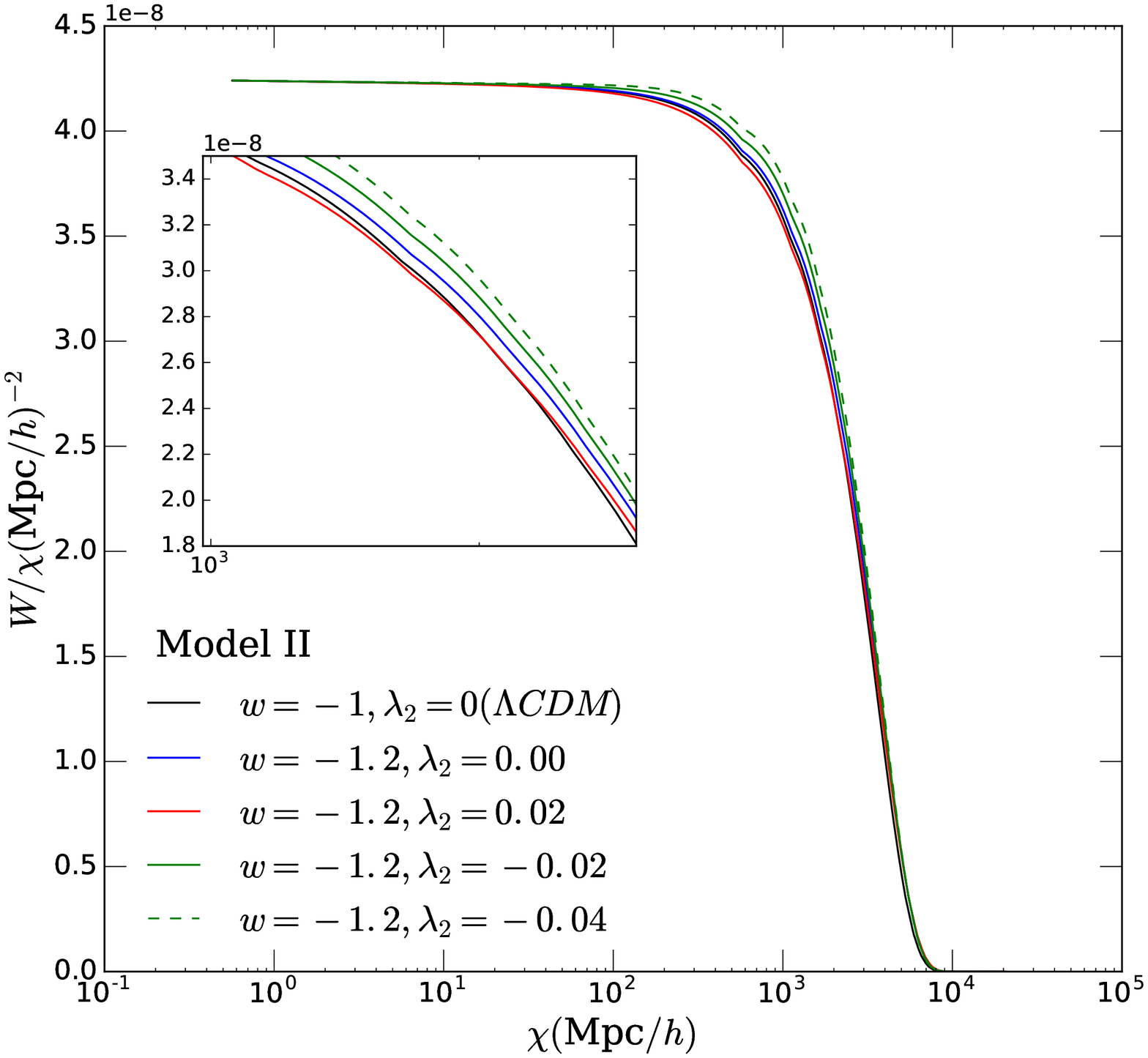}
\includegraphics[scale=0.3]{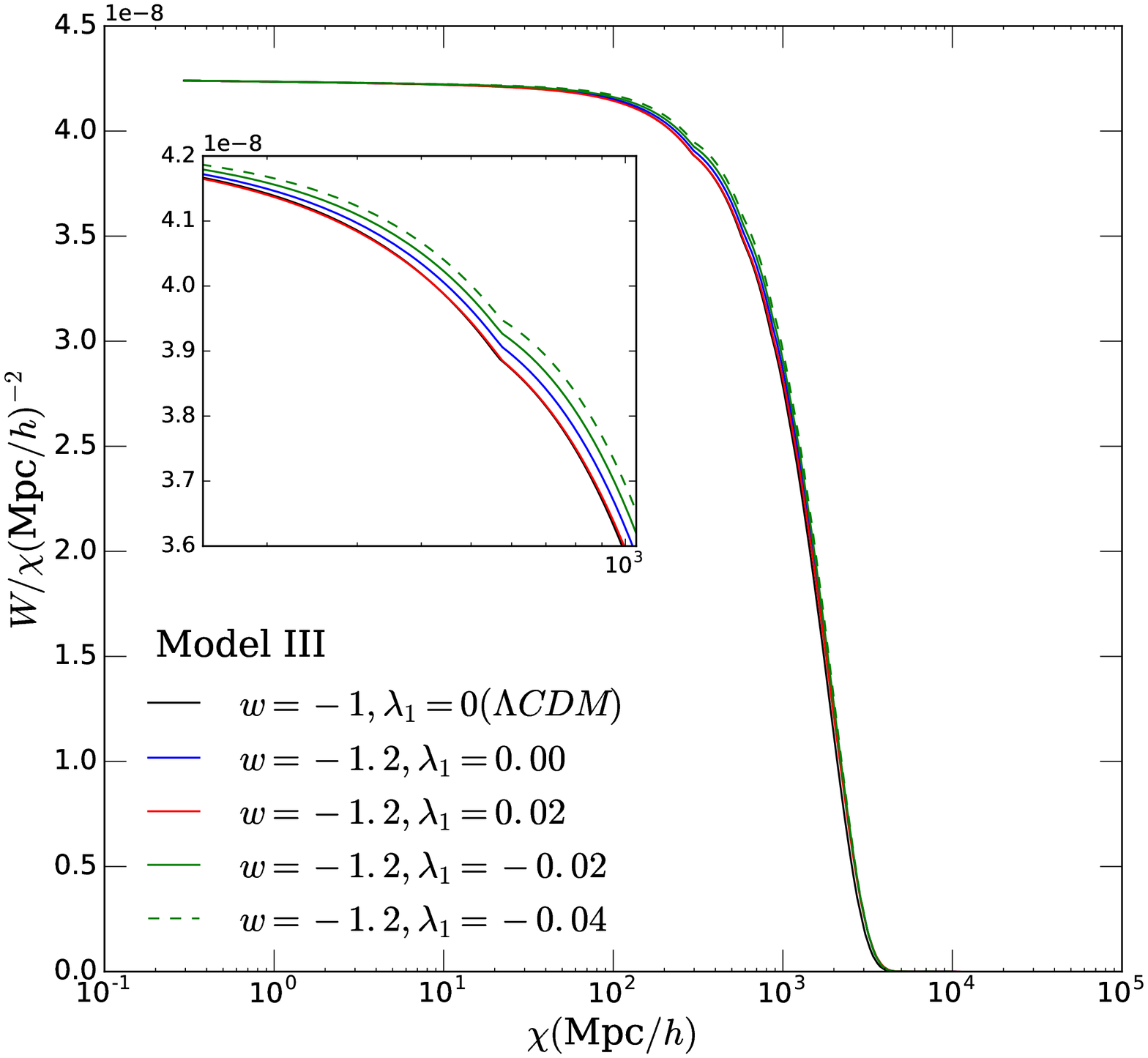}
\includegraphics[scale=0.3]{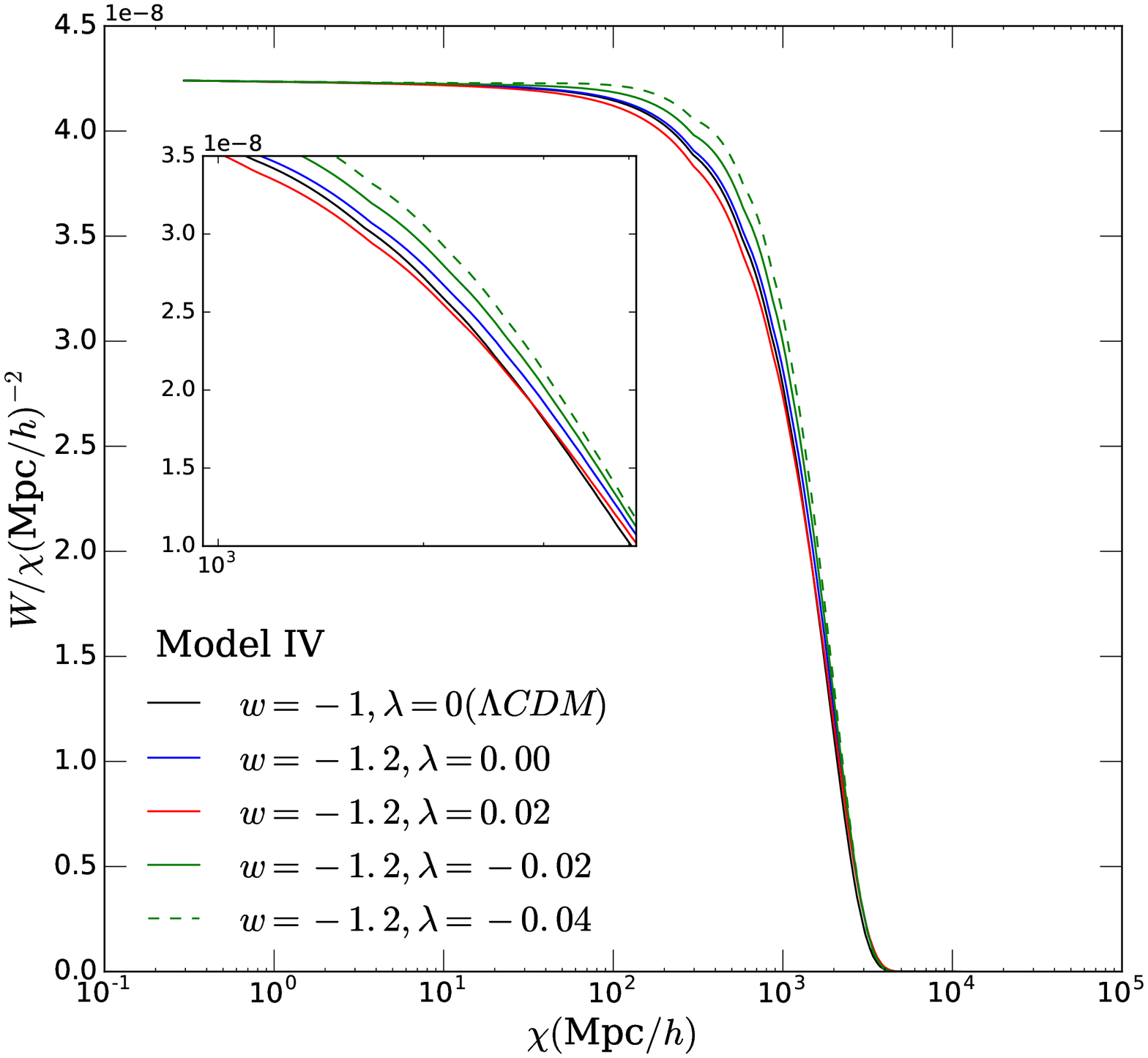}
\caption{Lensing efficiency function $W(\chi)/\chi$ for different dark energy interacting models with constant equation of state of dark energy.}
\label{fig.W}
\end{figure*}

At $z=0$, the integration with redshift distribution in Eq. (\ref{eq.W1}) is unity and $3a^2H^2\Omega_c/(2c^2)$ is equal to the value at present. If we choose $H_0=100h\ \rm{km}/\rm{Mpc}$, $\Omega_{c0}=0.25$, we find that $W(\chi)/\chi$ approaches to  $4.17\times 10^{-8} (\rm{Mpc}/h)^{-2}$ in the limit $z\rightarrow 0$. This explains why all the models converge to a constant value of lensing efficiency in Fig. (\ref{fig.W}) in the low redshift limit.

Fig. (\ref{fig.W}) shows the lensing weighting function $W(\chi)/\chi$ for four interacting models. For models with no interactions, the value of $W(\chi)/\chi$ is lower when a constant $w > -1$ at high redshift and it can be even lower when there is energy flow from dark energy to dark matter, but it gets higher if there are interactions that allow dark matter to decay into dark energy. For $w<-1$, the value of lensing weighting function can be enhanced, while the effect of the interaction remains the same qualitatively. Due to the interactions between dark sectors, a lower (higher) lensing weight than the prediction of $\Lambda$CDM can gradually become higher (lower). When both of them are equal at a characteristic comoving distance, we call it a `crossing'. For four interacting models, we find that the characteristic crossing of $W(\chi)/\chi$ is around 1000 $\rm{Mpc}/h$ which is determined by both the equation of state and the direction of the flow $Q$.

For a time evolving $w$, the lensing weighting function $W(\chi)/\chi$ behaves qualitatively the same as for the constant $w$. When $w>-1$, the values of the weighting function are always smaller than $\Lambda$CDM prediction, whereas the values are always larger if $w < -1$ \cite{Huterer2002}.

Future surveys will provide photometric redshift information for source galaxies. This enables us to make tomographic measurements by dividing the galaxy populations into redshift bins. The weighting function in a redshift bin $i$ is given by
\begin{equation}
\label{eq.Wi}
W^{(i)}(\chi)=\left\{
\begin{aligned}
&\frac{3}{2c^2}a^2H^2\Omega_{c}\chi\int_{\rm{max}(\chi,\chi_i)}^{\chi_{i+1}}d\chi'p(z)&\frac{dz}{d\chi'}\frac{\chi'-\chi}{\chi'},\\
&&\chi \le \chi_{i+1}\\
&0.&\chi>\chi_{i+1}
\end{aligned}
\right.
\end{equation}

\subsection{Convergence power spectrum}

\begin{figure*}
\includegraphics[scale=0.3]{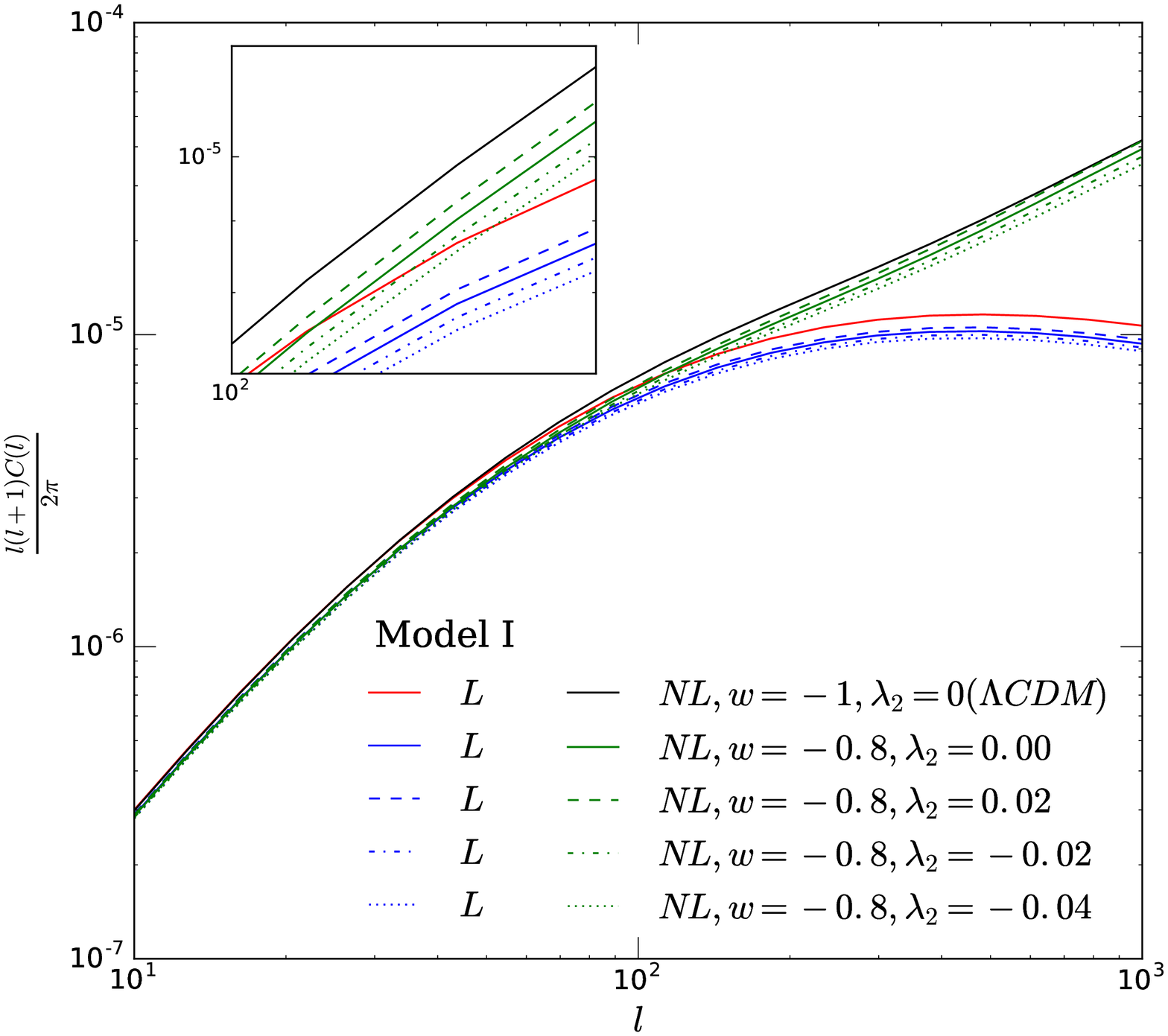}
\includegraphics[scale=0.3]{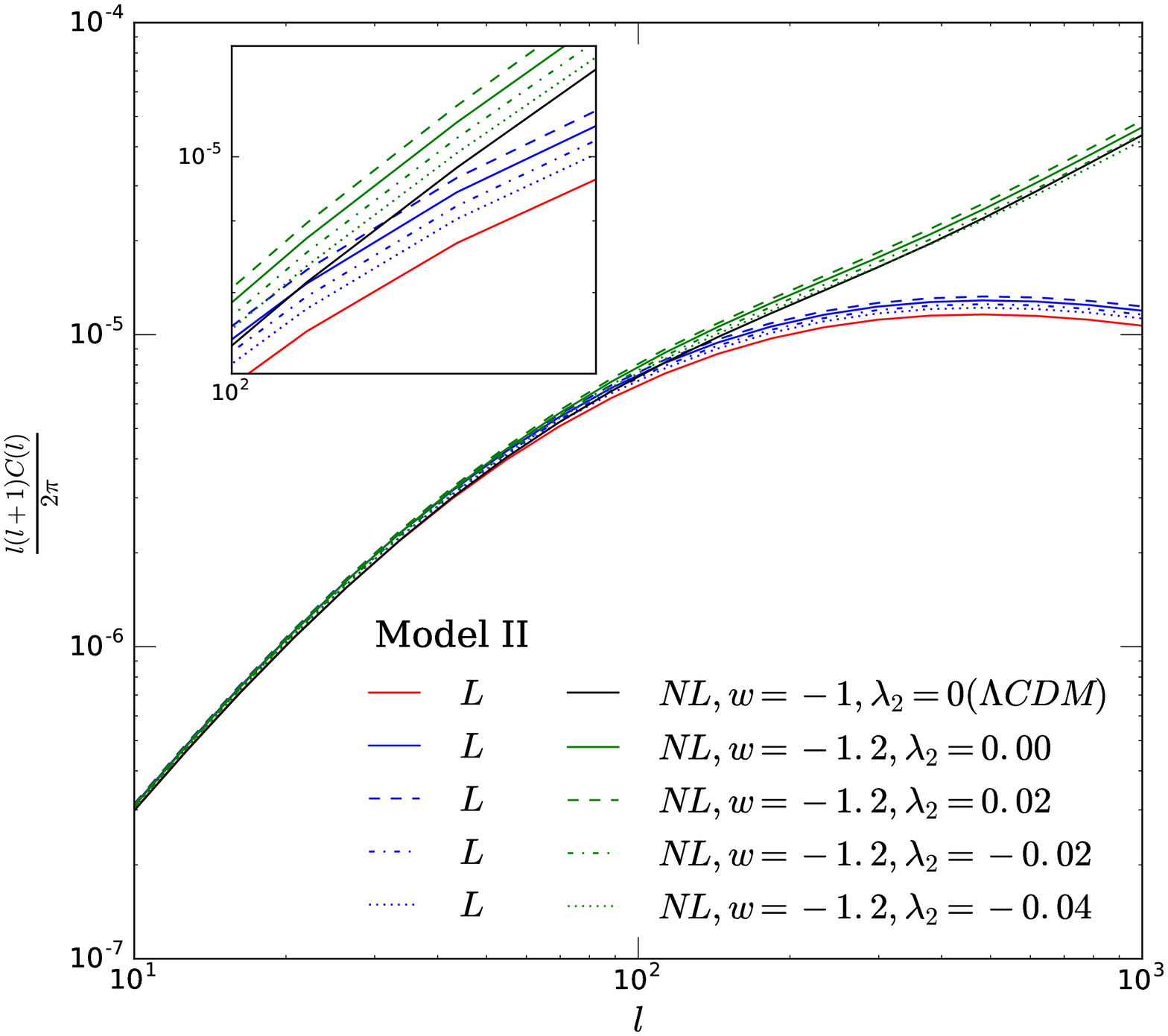}
\includegraphics[scale=0.3]{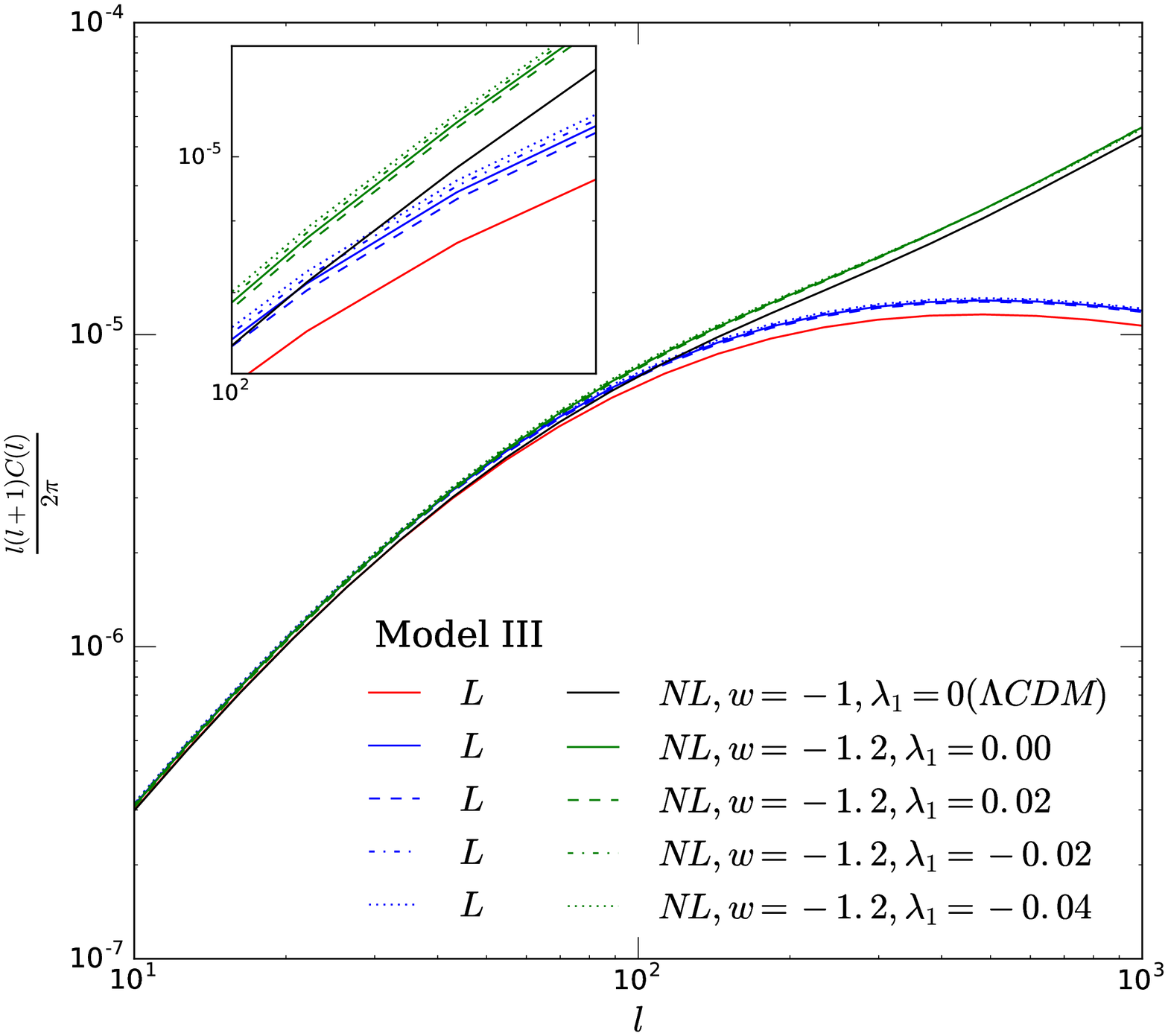}
\includegraphics[scale=0.3]{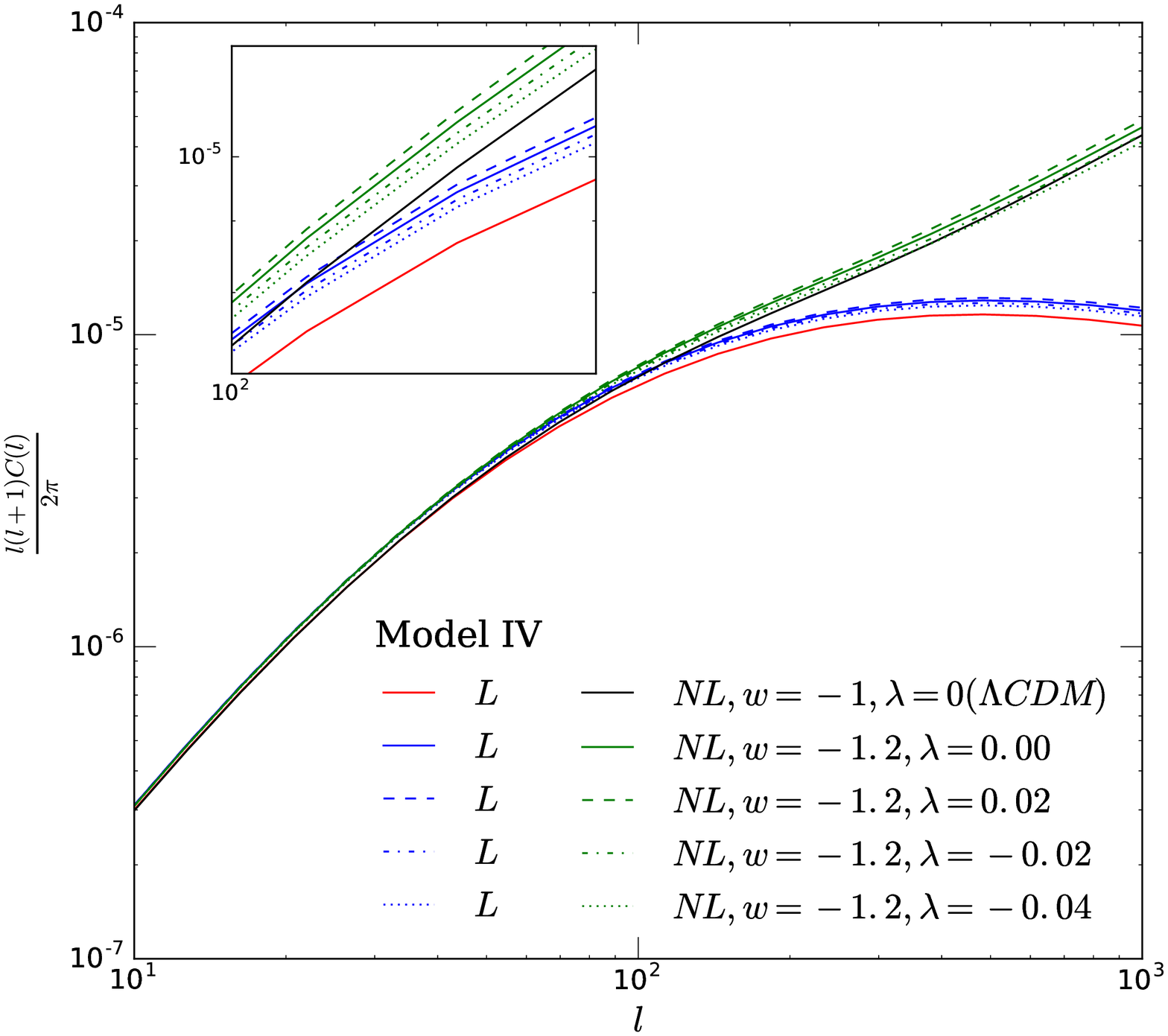}
\caption{The linear and nonlinear convergence power spectra for different interacting dark energy models with constant dark energy equation of state for a one-bin tomography model. In all the subfigures, `L' and `NL' refer to linear and non-linear power spectra.}
\label{fig.Cl_l_nl}
\end{figure*}

The lensing convergence field can be decomposed into spherical harmonics with coefficients $\kappa_{lm}$ ( = $\int d\bm{\theta}\kappa(\bm{\theta})Y_{lm}^*$). The convergence power spectrum is then defined by \cite{Bartelmann2001}
\begin{equation}
\label{eq.Pdefine}
\langle \kappa_{l_1m_1}\kappa_{l_2m_2}\rangle=\delta_{l_1l_2}\delta_{m_1m_2}C_{l_1}.
\end{equation}
According to the Limber approximation \cite{Limber1954,Kaiser1992}, the convergence power spectrum $C^{(ij)}_l$ between redshift bins $i$ and $j$ can be written as
\begin{equation}
\label{eq.Cl}
C^{(ij)}_l=\int_{0}^{\chi_H}d\chi W^{(i)}(\chi)W^{(j)}(\chi)\chi^{-2}P_{\delta}\left(\frac{l}{\chi},\chi\right),
\end{equation}
where $P_{\delta}$ is a non-linear matter power spectrum. Cosmological N-body simulations can be used to derive the nonlinear power spectrum precisely, but a numerical implementation of a general interacting dark matter and dark energy model has not been made and is certainly beyond the scope of this work. Approximately, an estimate of the non-linear power spectrum can be obtained if we assume that the nonlinear evolution of matter perturbations is weakly dependent of cosmological parameters. Under such an assumption, the non-linear matter power spectrum can be simply parametrized by a Halofit model. There has been a plenty of applications of the Halofit model to forecasting cosmological parameters in the literature, such as the constraints on  coupled dark energy model\cite{Vacca2008}, the bispectrum constrains on the neutrino masses \cite{Namikawa2016}, the cosmological forecasts from weak gravitational lensing surveys \cite{Harrison2016}. We follow Ref. \cite{Takahashi2012} to make corrections to the linear power spectrum for the interacting models using the Halofit method \cite{Simth2003}. Fig. (\ref{fig.Cl_l_nl}) shows the significant power excess of nonlinear clustering at small scales for a one-bin tomography model.

As Eq. (\ref{eq.Cl}) indicates, the convergence power spectrum consists of both the lensing weighting function $W(\chi)$ and the matter power spectrum $P_{\delta}$. The lensing weighting function largely relies on the background evolution that can be changed by the parameter $w$ and the interaction parameters. The non-linear matter power spectrum which is mainly affected by the perturbations of the dark sectors can be parametrized by a Halofit model in which $w$ is an independent parameter and the fraction of dark energy density $\Omega_c(z)$ is required in and dependent of the interaction parameters. All of the five models we consider in this work have different features in both the background evolution and matter perturbations. From the calculations, we find that the convergence power spectra with models III and V obtain most of the parameter dependencies from the weighting functions, but for models I, II and IV, the parameter dependencies mainly come from the non-linear power spectra. All of these parameter dependencies make the convergence power spectrum an ideal probe to constraining the interaction models.

\subsection{Bispectrum and its covariance}

The lensing bispectrum among the convergence fields in redshift bins $i$, $j$ and $k$ is defined as
\begin{equation}
\label{eq.Bdefine}
\langle \kappa^{(i)}_{l_1m_1}\kappa^{(j)}_{l_2m_2}\kappa^{(k)}_{l_3m_3}\rangle=\left(
\begin{matrix}
l_1&l_2&l_3\\
m_1&m_2&m_3
\end{matrix}
\right)B_{l_1l_2l_3}^{(ijk)},
\end{equation}
where $\kappa_{lm}$s are the spherical harmonic coefficients of the convergence field. The triangle condition $|l_i-l_j|\leq l_k\leq l_i+l_j$ is automatically satisfied by the 3-j Wigner symbol.

The bispectrum $B^{(ijk)}_{l_1l_2l_3}$ can be approximately related to a flat-sky bispectrum $B_{ijk}(\bm{l}_1,\bm{l}_2,\bm{l}_3)$ by \cite{Takada2004}
\begin{eqnarray}
\label{eq.Bfull}
B^{(ijk)}_{l_1l_2l_3}&\simeq& \left(
\begin{matrix}
l_1&l_2&l_3\\
0&0&0
\end{matrix}
\right)\sqrt{\frac{(2l_1+1)(2l_2+1)(2l_3+2)}{4\pi}}\nonumber\\
&\times&B_{ijk}(\bm{l}_1,\bm{l}_2, \bm{l}_3).
\end{eqnarray}

Using the Limber approximation, the flat-sky bispectrum is expressed as
\begin{eqnarray}
\label{eq.Bflat}
B_{(ijk)}(\bm{l}_1,\bm{l}_2,\bm{l}_2)&=&\int_{0}^{\chi_H}W^{(i)}(\chi)W^{(j)}(\chi)W^{(k)}(\chi)\chi^{-4}\nonumber\\
&\times&B_{\delta}(\bm{k}_1,\bm{k}_2,\bm{k}_3),
\end{eqnarray}
where $\bm{k}_p=\bm{l}_p/\chi$, $p=1,2,3$. Here $B_{\delta}(\bm{k}_1,\bm{k}_2,\bm{k}_3)$ is a three-dimensional matter bispectrum expressed as \cite{Fry1984}
\begin{eqnarray}
\label{eq.Bdelta}
B_{\delta}(\bm{k}_1,\bm{k}_2,\bm{k}_3)&=&2F_2(\bm{k}_1,\bm{k}_2)P_{\delta}^{NL}(\bm{k}_1,z)P_{\delta}^{NL}(\bm{k}_2,z)\nonumber\\
&+&2\, \rm{perms}.
\end{eqnarray}
with the effective kernel $F_2(\bm{k}_1,\bm{k}_2)$ in hyper-extended perturbation theory \cite{Scoccimarro2001}
\begin{equation}
\label{eq.F2}
\begin{split}
F_2(\bm{k}_1,\bm{k}_2)=&\frac{5}{7}a(n,k_1)a(n,k_2)+\\
&\frac{1}{2}\left(\frac{k_1}{k_2}+\frac{k_2}{k_1}\right)\frac{\bm{k}_1\cdot \bm{k}_2}{k_1k_2}b(n,k_1)b(n,k_2)+\\
&\frac{2}{7}\left(\frac{\bm{k}_1\cdot \bm{k}_2}{k_1k_2}\right)^2c(n,k_1)c(n,k_2),
\end{split}
\end{equation}
where the fitting functions $a(n,k)$, $b(n,k)$ and $c(n,k)$ are given by Ref \cite{Gil-Marin2012} which are extensions to formula in Ref \cite{Scoccimarro2001}. These functions are expressed as
\begin{gather}
\label{eq.a}
a(n,k)=\frac{1+\sigma_{8}^{a_6}(z)[0.7Q_3(n)]^{1/2}(qa_1)^{n+a_2}}{1+(qa_1)^{n+a_2}},\\
\label{eq.b}
b(n,k)=\frac{1+0.2a_3(n+3)(qa_7)^{n+3+a_8}}{1+q^{n+3.5+a_8}},\\
\label{eq.c}
c(n,k)=\frac{1+4.5a_4/[1.5+(n+3)^4](qa_5)^{n+3+a_9}}{1+(qa_5)^{n+3.5+a_9}},
\end{gather}
with the best fit parameters $a_1=0.484$, $a_2=3.740$, $a_3=-0.849$, $a_4=0.392$, $a_5=1.013$, $a_6=-0.575$, $a_7=0.128$, $a_8=-0.722$ and $a_9=-0.926$. The parameter $\sigma_{8}(z)$ denotes the variance of the matter density fluctuations smoothed by a top-hat window with radius $8\rm{Mpc}/h$ at an arbitrary redshift and it can be derived from the current value via $\sigma_8(z)=D(z)\sigma_8$. The function $Q_3(n)=(4-2^n)/(1+2^{n+1})$, and $n$ is the slope of the linear matter power spectrum, i.e., $n(k)=\text{d}\ln P_{\delta}^L/\text{d}\ln k$. The quantity $q=k/k_{\rm{NL}}$ is rescaled by the nonlinear wave number $k_{\rm{NL}}$ satisfying $(k_{\rm{NL}}^{3}/2\pi^3)P_{\delta}^{L}(k_{\rm{NL}})=1$.

\begin{figure*}
\includegraphics[scale=0.3]{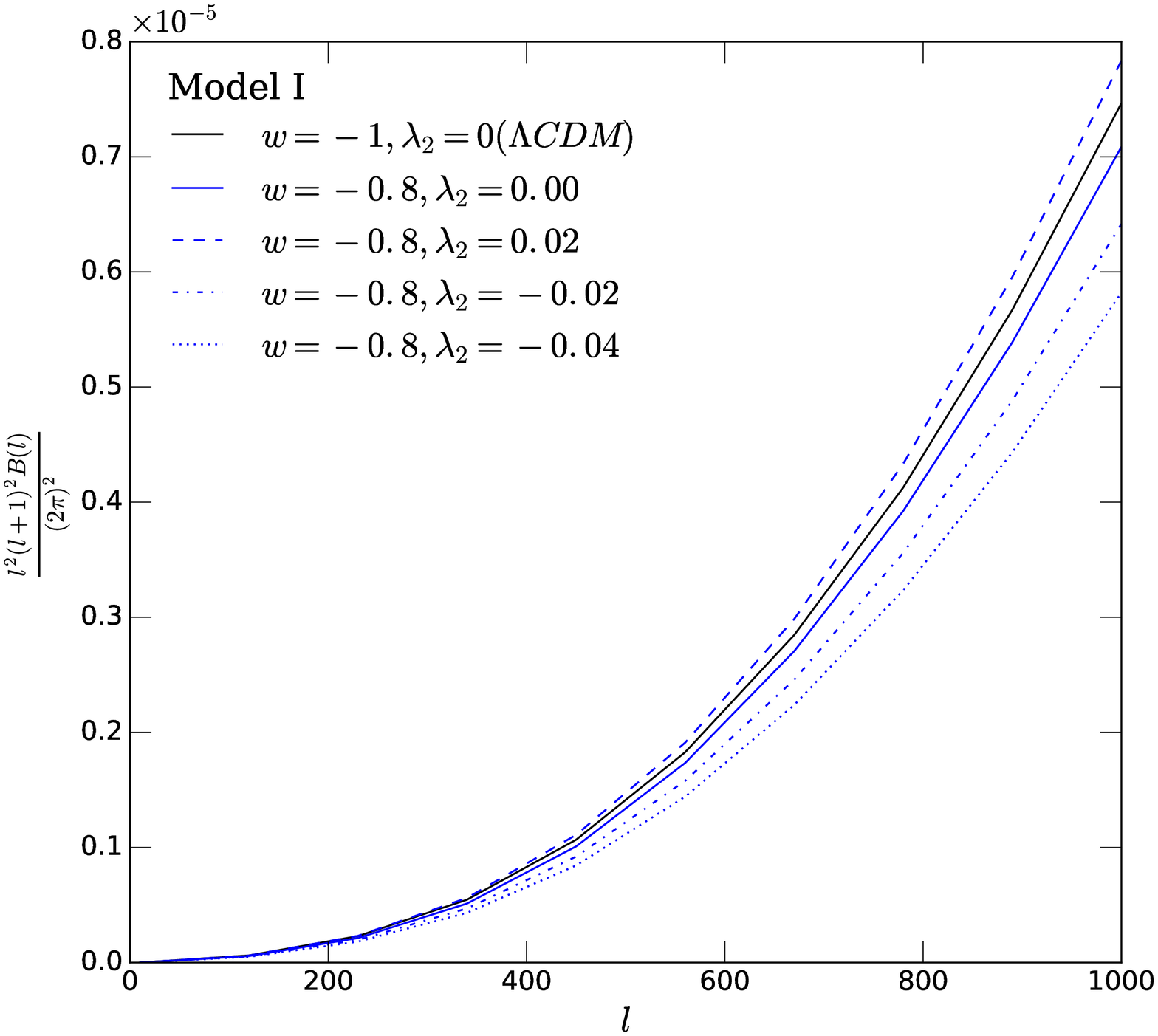}
\includegraphics[scale=0.3]{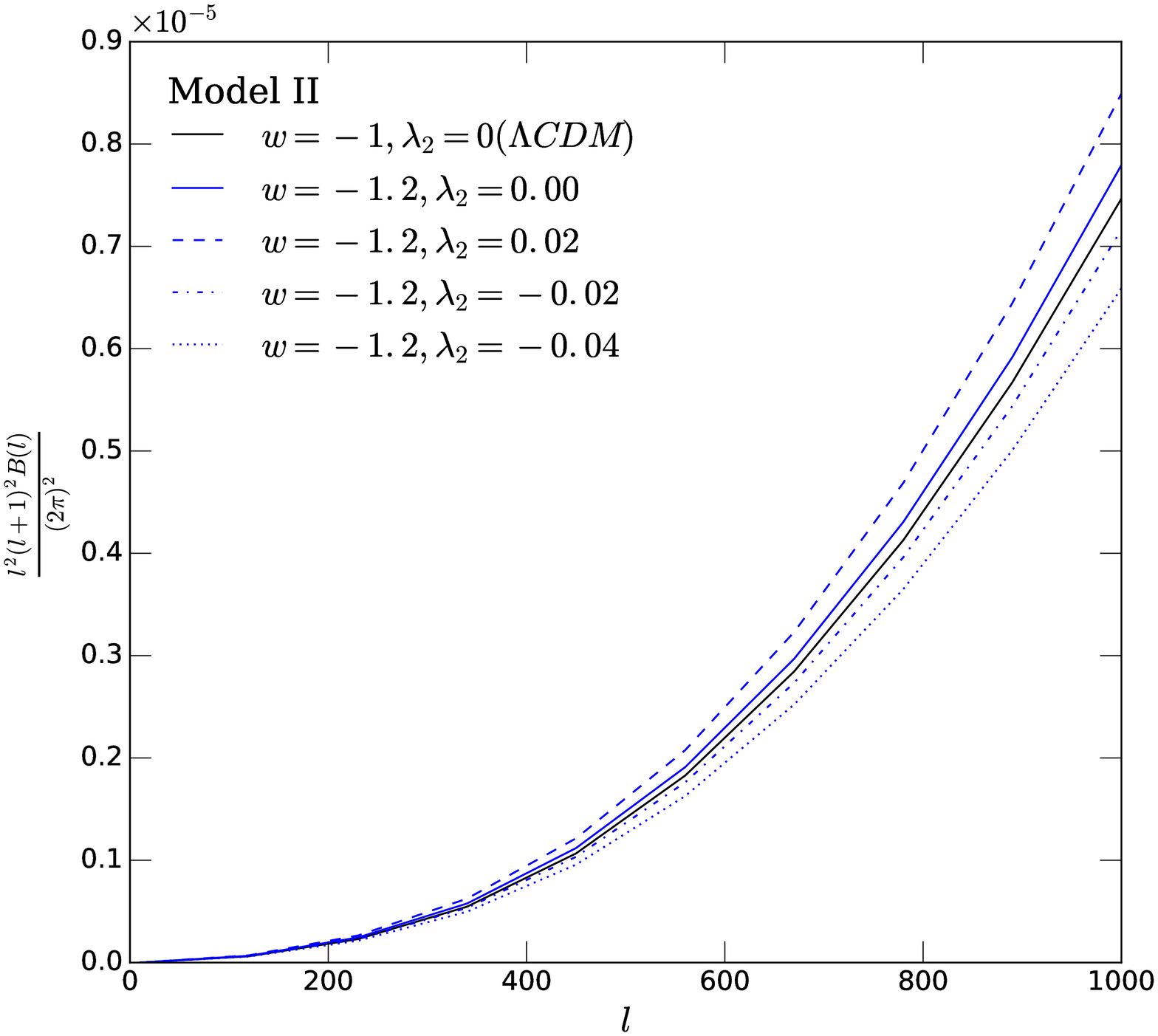}
\includegraphics[scale=0.3]{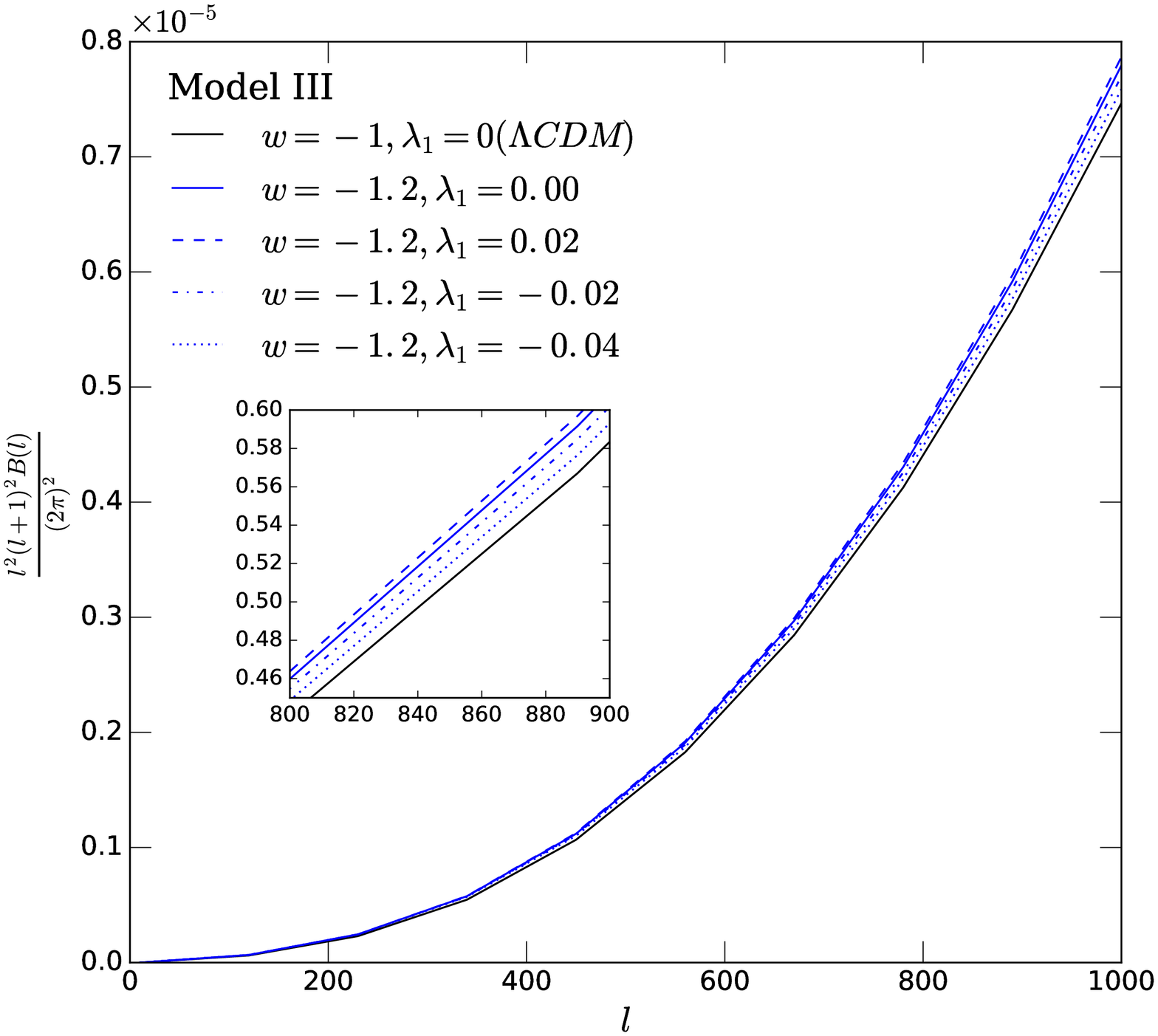}
\includegraphics[scale=0.3]{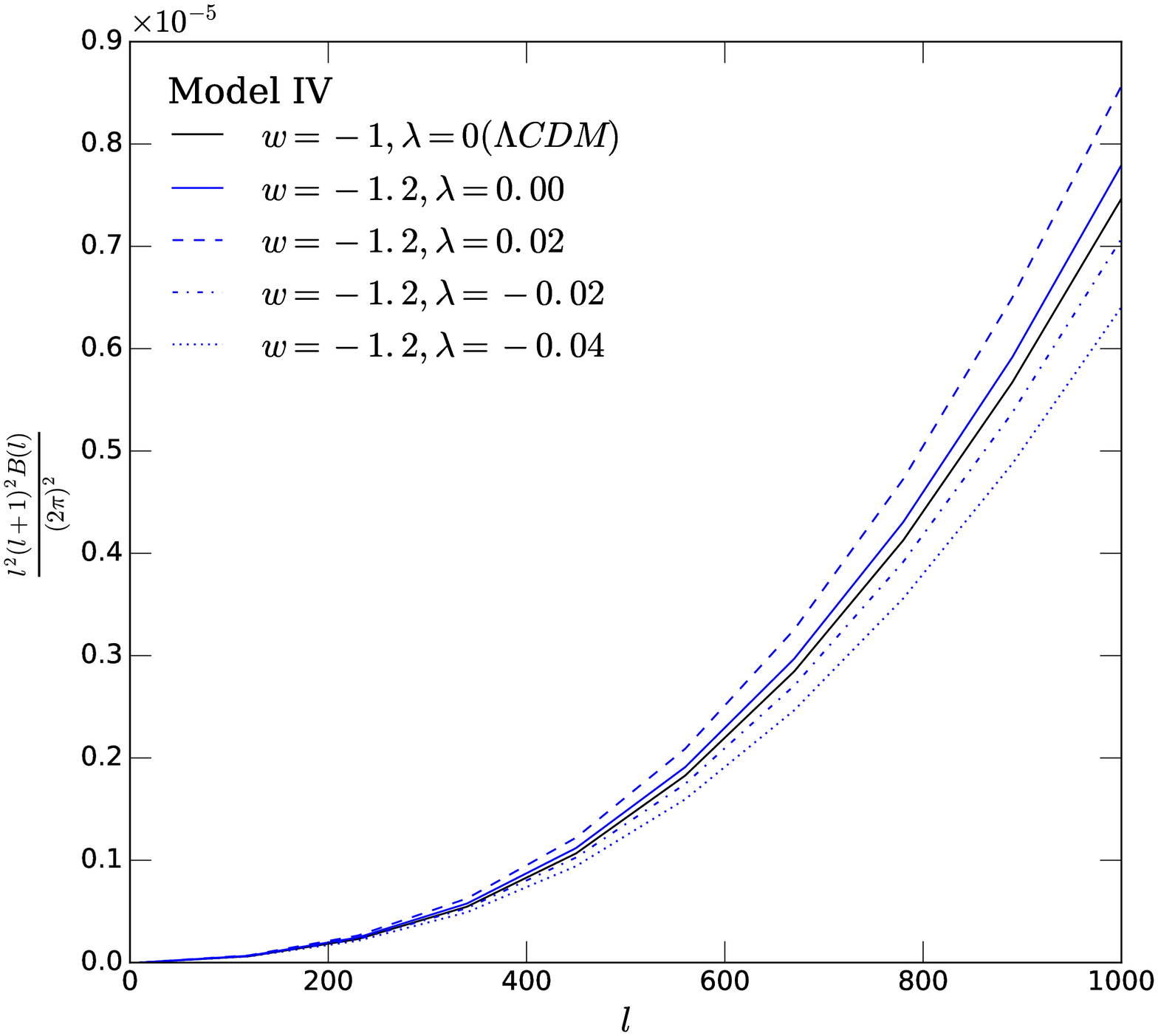}
\caption{Equilateral convergence bispectra $B(l)$ with $l_1=l_2=l_3=l$ for different interacting dark energy models with constant dark energy equation of state for a one-bin tomography model.}
\label{fig.Bl_e}
\end{figure*}

Similar to the weak lensing power spectrum, bispectrum is also determined by the lensing weighting function and the non-linear matter power spectrum. The main difference lies in the powers of both $W(\chi)$ and $P_{\delta}$, i.e., $W(\chi)^2P_{\delta}$ for power spectrum but $W(\chi)^3P_{\delta}^2$ for bispectrum. All the lensing bispectra with equilateral configurations for different models are shown in Fig. (\ref{fig.Bl_e}). It is seen that the bispectrum is more sensitive to the interaction between dark energy and dark matter than the power spectrum. When there is no interaction, the equation of state has a great impact on the lensing weighting functions of the bispectrum. But the matter power spectrum begins to dominate the bispectrum for all the models when there is an interaction between dark sectors. These two factors would result in larger differences in the bispectrum than the power spectrum.

The bispectrum covariance matrix for tomography is approximated by \cite{Takada2004}
\begin{eqnarray}
\label{eq.CovB}
&&\text{Cov}[B^{(ijk)}_{l_1,l_2,l_3},B^{(i'j'k')}_{l_1,l_2,l_3}]\nonumber\\
&=&\bigtriangleup(l_1,l_2,l_3)f_{\rm{sky}}^{-1}\hat C^{(ii')}_{l_1}\hat C^{(jj')}_{l_2}\hat C^{(kk')}_{l_3},
\end{eqnarray}
where $\bigtriangleup(l_1,l_2,l_3)=1$ if $l_1\neq l_2\neq l_3$, $\bigtriangleup(l_1,l_2,l_3)=2$ if $l_i=l_j, i\neq j$ and $\bigtriangleup(l_1,l_2,l_3)=6$ if $l_1=l_2=l_3$. The parameter $f_{\rm{sky}}$ is the sky fraction, which is about $0.5$ in our toy model. The quantity $\hat C^{(ij)}_l$ is the observed power spectrum \cite{Hu1999} between redshift bins $i$ and $j$ with a shot noise included, i.e.,
\begin{equation}
\label{eq.Cobs}
\hat C^{(ij)}_l=C^{(ij)}_l+\delta_{ij}\frac{\sigma_\epsilon}{n_i},
\end{equation}
where $\sigma_\epsilon$ is the intrinsic ellipticity dispersion of the galaxy sample, and the average number density of galaxies in the $i$-th redshift bin $n_i$ is given by
\begin{equation}
\label{eq.ni}
n_i=\int_{\chi_i}^{\chi_{i+1}}d\chi'p(z)\frac{dz}{d\chi'}.
\end{equation}

Here we only consider the Gaussian contribution of the covariance matrix. In Ref \cite{Kayo2013}, Figure 5 shows that more than half of the covariance comes from non-Gaussian terms at $\ell\sim 2000$ for equilateral triangles. Ref \cite{Sato2013} has concluded that although the non-Gaussian convariance can reduce the overal signal-to-noise by a factor of 3 at $\ell=2000$, it can only degrade parameter constraints by 15\% if a shot noise is taken into account. For this work, our shot noise model of a DUNE--type weak lensing experiment is even higher than the HSC--type model adopted in their work so the impact of the non-Gaussian covariance on the parameters, especially $\lambda_1$ and $\lambda_2$ in our model, should be even smaller than 15\%. Therefore, for our Fisher matrix analysis, it is a reasonable assumption to ignore the non-Gaussian covariance, while only keeping the Gaussian part that substantially simplifies the forecasting.

\section{Fisher matrix analysis}

In this section, we adopt the Fisher matrix formalism \cite{Tegmark1997} to estimate uncertainties of the cosmological parameters $\Omega_{c0}$, $\sigma_8$, $h$, the parameters of the equation of state $w_0$, $w_a$, and the coupling parameters $\lambda_1$, $\lambda_2$. Here we assume a fiducial weak lensing survey with specifications the same as the proposed DUNE project \cite{Refregier2006}. Table.\ref{DUNE} lists all the parameters of DUNE.

\begin{table}
\caption{Parameters of the weak lensing survey of DUNE. \label{DUNE}}
\begin{tabular}{l|l}
\hline
galaxy number per steradian  & $n=4.7\times 10^8$\\
solid angle & $\bigtriangleup\Omega=2\pi$\\
sky fraction & $f_{\rm{sky}}=1/2$\\
survey depth & $z_0=0.64$\\
intrinsic ellipticity dispersion &$\sigma_{\epsilon}=0.3$\\
\hline
\end{tabular}
\end{table}

In principle, both the convergence power spectrum and bispectrum can be used to constrain the interaction models but their sensitivities may differ. We have verified that the bispectrum with tomography has better signal-to-noise ratios and tighter bounds on the cosmological parameters than the convergence power spectrum. As seen in Table. \ref{tab.PvsB}, the error bars from the convergence power spectrum without tomography could be four times larger than that of the bispectrum. Therefore, we choose the bispectrum to constrain the cosmological parameters, particularly, the interaction parameters. Furthermore, the tomography will add extra information from the line of sight so the bounds could be reduced even further. From Table. \ref{tab.B1vsB2}, it is seen that the error bars from bispectrum with no tomography could be two to three times larger than the ones from bispectrum with a two-bin tomography. 

\begin{table}

\caption{Comparisons between the power spectrum and the bispectrum without any tomography. \label{tab.PvsB}}
\resizebox{\linewidth}{!}{%
\begin{tabular}{l|l|l}
\hline
Fiducial Model($\Lambda$CDM) & Power Spectrum & Bispectrum\\
\hline
$\Omega_c=0.25$ &$\bigtriangleup \Omega_c=0.017674$\ \ \ \ \ \ \ \ &$\bigtriangleup \Omega_c=0.010537$\ \ \ \ \ \ \ \ \\
$\sigma_8=0.8$ & $\bigtriangleup \sigma_8=0.030334$ & $\bigtriangleup \sigma_8=0.023986$\\
$w_0=-1$ & $\bigtriangleup w_0=0.748681$ & $\bigtriangleup w_0=0.182274$ \\
$\lambda_1=0$ & $\bigtriangleup \lambda_1=0.119159$ & $\bigtriangleup \lambda_1=0.093232$ \\
$\lambda_2=0$ & $\bigtriangleup \lambda_2=0.038550$ & $\bigtriangleup \lambda_2=0.037720$  \\
$h=0.72$ & $\bigtriangleup h=0.481874$ & $\bigtriangleup h=0.236921$ \\
\hline
\end{tabular}}
\end{table}

\begin{table}

\caption{Comparisons between the bispectrum with no tomography and 2-bin tomography. \label{tab.B1vsB2}}
\resizebox{\linewidth}{!}{%
\begin{tabular}{l|l|l}
\hline
Fiducial Model($\Lambda$CDM) & No Tomography & 2-Bin Tomography\\
\hline
$\Omega_c=0.25$ &$\bigtriangleup \Omega_c=0.010537$\ \ \ \ \ \ \ \ &$\bigtriangleup \Omega_c=0.007350$\ \ \ \ \ \ \ \ \\
$\sigma_8=0.8$ & $\bigtriangleup \sigma_8=0.023986$ & $\bigtriangleup \sigma_8=0.013507$\\
$w_0=-1$ & $\bigtriangleup w_0=0.182274$ & $\bigtriangleup w_0=0.094660$ \\
$\lambda_1=0$ & $\bigtriangleup \lambda_1=0.093232$ & $\bigtriangleup \lambda_1=0.045974$ \\
$\lambda_2=0$ & $\bigtriangleup \lambda_2=0.037720$ & $\bigtriangleup \lambda_2=0.010776$  \\
$h=0.72$ & $\bigtriangleup h=0.236921$ & $\bigtriangleup h=0.141352$ \\
\hline
\end{tabular}}
\end{table}

\begin{table*}
\caption{Cosmological implications from hypothetical models with interaction parameters being the upper bounds.}\label{cosmo_imp}
\begin{tabular}{c|c|c|c|c|c}
\hline    
\rm{Models}&\rm{$\Lambda$CDM} (\rm{default})&$\lambda_1=0.07$&$\lambda_1=-0.07$&$\lambda_2=0.02$&$\lambda_2=-0.02$\\
\hline
\ \ \ \rm{Age} (\rm{Gyr})\ \ \ &13.777 &14.362 (4.24\%)&13.215 (4.07\%)&13.937 (1.16\%)&13.630 (1.06\%)\\
\hline
$C^{\kappa\kappa}_{\ell=1000}$&$2.737\times 10^{-10}$&$2.731\times 10^{-10}$ (0.19\%)&$2.684\times 10^{-10}$ (1.91\%)&$2.889\times 10^{-10}$ (5.56\%)&$2.596\times 10^{-10}$ (5.14\%)\\
\hline
$B^{\kappa\kappa\kappa}_{\ell=1000}$&$2.941\times 10^{-16}$&$2.986\times 10^{-16}$ (1.53\%)&$2.802\times 10^{-16}$ (4.74\%)&$3.227\times 10^{-16}$ (9.70\%)&$2.684\times 10^{-16}$ (8.73\%)\\
\hline
\end{tabular}
\end{table*}

The Fisher matrix can be expressed as
\begin{eqnarray}
\label{eq.Fbis}
F_{\alpha \beta}&=&\sum_{l_1,l_2,l_3=l_{\rm{min}}}^{l_{\rm{max}}}\sum_{(ijk),(i'j'k')}\nonumber\\
&&\frac{\partial B^{(ijk)}_{l_1,l_2,l_3}}{\partial x_{\alpha}}[\text{Cov}[B^{(ijk)}_{l_1,l_2,l_3},B^{(i'j'k')}_{l_1,l_2,l_3}]]^{-1}\nonumber\\
&\times&\frac{\partial B^{(i'j'k')}_{l_1,l_2,l_3}}{\partial x_{\beta}},
\end{eqnarray}
with the condition $l_1\leq l_2\leq l_3$ so that every triangle configuration is only counted once.

\begin{figure*}
\centering
\includegraphics[scale=0.6]{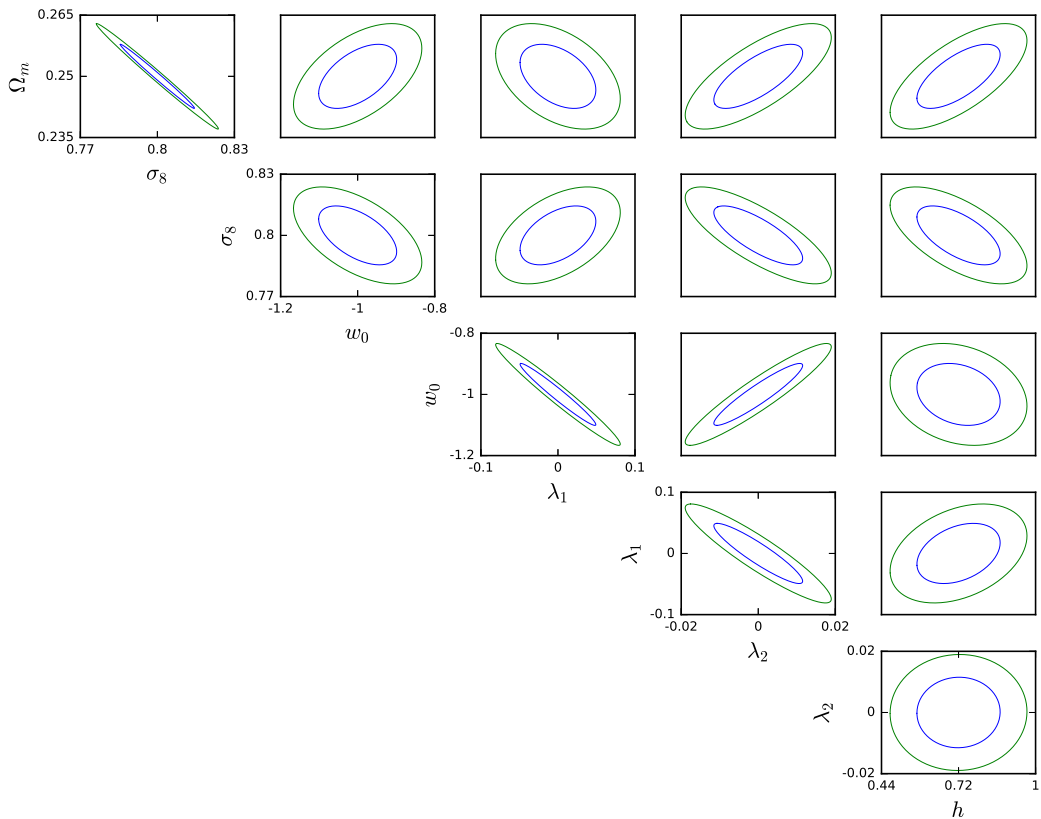}
\caption{Parameter constraints from the Fisher matrix analysis with constant dark energy equation of state for a two-bin tomography model. The blue and green ellipses correspond to $1\sigma$ and $2\sigma$. }
\label{fig.Fisher}
\end{figure*}

\begin{figure*}
\centering
\includegraphics[scale=0.6]{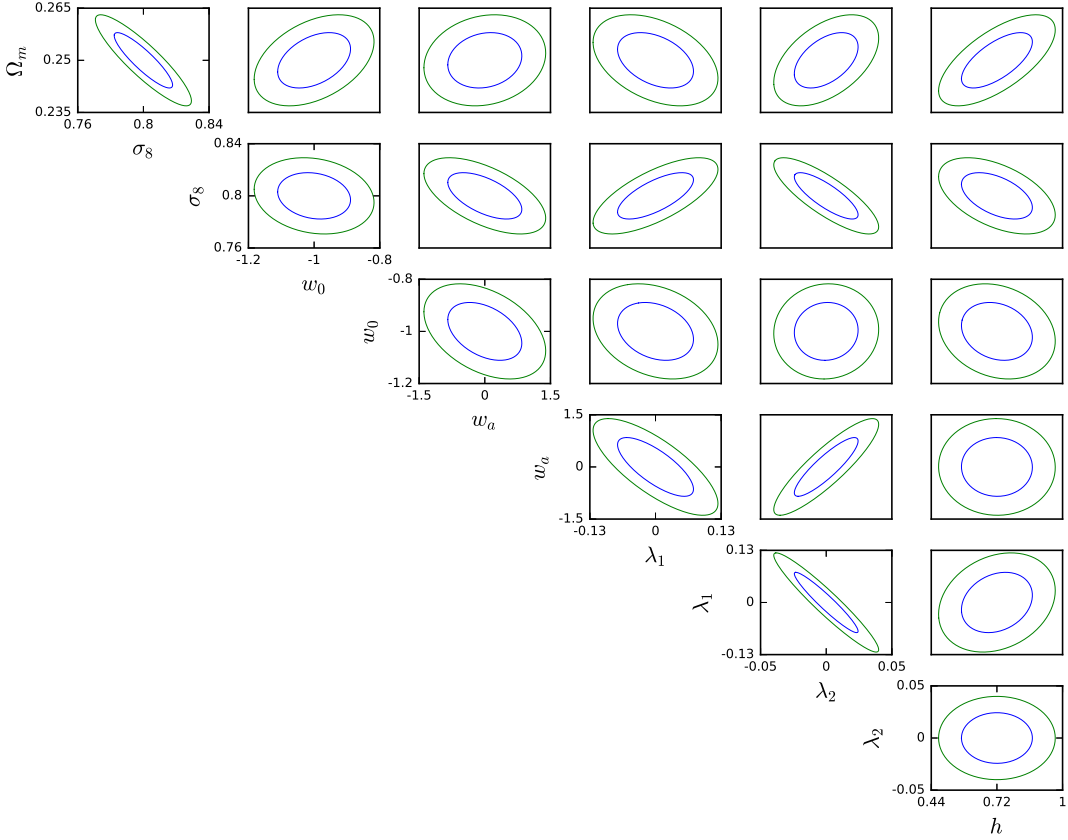}
\caption{Parameter constraints from the Fisher matrix analysis with time evolving dark energy equation of state $w=w_0+w_a(1-a)$ for a two-bin tomography model. The blue and green ellipses correspond to $1\sigma$ and $2\sigma$.}
\label{fig.Fisherwa}
\end{figure*}

\begin{table}

\caption{Expected accuracy on the parameters from two-bin weak lensing tomography. \label{tab.1sigma}}
\resizebox{\linewidth}{!}{%
\begin{tabular}{l|l|l}
\hline
Fiducial Model($\Lambda$CDM) & $w=w_0$ & $w=w_0+w_a(1-a)$\\
\hline
$\Omega_c=0.25$ &$\bigtriangleup \Omega_c=0.007350$\ \ \ \ \ \ \ \ &$\bigtriangleup \Omega_c=0.007446$\ \ \ \ \ \ \ \ \\
$\sigma_8=0.8$ & $\bigtriangleup \sigma_8=0.013507$ & $\bigtriangleup \sigma_8=0.016626$\\
$w_0=-1$ & $\bigtriangleup w_0=0.094660$ & $\bigtriangleup w_0=0.103739$ \\
$w_a=0$ & $\bigtriangleup w_a=0$ & $\bigtriangleup w_a=0.790401$ \\
$\lambda_1=0$ & $\bigtriangleup \lambda_1=0.045974$ & $\bigtriangleup \lambda_1=0.070270$ \\
$\lambda_2=0$ & $\bigtriangleup \lambda_2=0.010776$ & $\bigtriangleup \lambda_2=0.022699$  \\
$h=0.72$ & $\bigtriangleup h=0.141352$ & $\bigtriangleup h=0.141358$ \\
\hline
\end{tabular}}
\end{table}

We consider a two-bin weak lensing tomography and present results of the Fisher matrix analysis. The multipole range is limited to $10^2 < l < 2\times10^3$ and a larger $l_{\rm{max}}$ will not significantly affect the results \cite{Vacca2008}. Two redshift bins $0<z<0.9$ and $z>0.9$ are chosen to have equal number of galaxy density. The fiducial cosmological parameters for the $\Lambda$CDM model is $\{\Omega_{c0}, \sigma_8, w_0, w_a, \lambda_1, \lambda_2, h\}$ = $\{0.25, 0.8, -1, 0, 0, 0, 0.72\}$. The Fisher matrix analysis is applied to models with two forms of equation of state: constant vs time evolving. The parameter sets for both cases are $\bm{P}$ = $\{\Omega_{c0},\sigma_8,w_0,\lambda_1,\lambda_2,h\}$ and $\bm{P}$ = $\{\Omega_{c0},\sigma_8,w_0,w_a,\lambda_1,\lambda_2,h\}$, respectively.

Table.\ref{tab.1sigma} lists the estimated errors on all the parameters. The margin of error is under $3\%$ for $\Omega_{c0}$ and $2\%$ for $\sigma_8$, whereas the error on $h$ is up to $20\%$. For the parameters of the equation of state, $w_0$ can be constrained as $\Delta w_0\simeq 0.1$ in the constant scenario. If $w$ evolves with time, the error on $w_0$ is in the same order as the constant $w$ while the constraint on $w_a$, $\Delta w_a\simeq 0.8$, is almost an order of magnitude larger than $w_0$. The errors of the coupling parameters are $\Delta\lambda_1\simeq 0.07$ and $\Delta\lambda_2\simeq 0.02$ and the constraints can be further tightened if the equation of state is set to be a constant. These results show the great potential of bispectrum tomography measurements in testing the interacting dark energy models.

The upper bounds $\Delta\lambda_1\simeq 0.07$ and $\Delta\lambda_2\simeq 0.02$ (68\% confidence level) only mean the maximum possible deviations from $\lambda_1=0$ and $\lambda_2=0$ ($\Lambda$CDM) for a given experiment that has a certain noise contamination. These upper bounds could be lower or higher, depending on the instrumental noise. For the DUNE--type experiment as we consider in this work, the upper bounds tell us that chances for being higher than 0.07 and 0.02 are statistically less than 32\%. The values of these bounds are indicative of figure-of-merit of a futuristic experiment.

Hypothetically, we can assume two cosmological models which describe interactions as strong as the upper bounds, i.e., $\lambda_1=0.07$ and $\lambda_2=0.02$, respectively. Also, we adopt values of the dark matter and dark energy fractions at present, i.e., $\Omega_c=8\pi G\rho_c/(3H^2)=0.25$ and $\Omega_d=8\pi G\rho_d/(3H^2)=0.75$. Given these numbers, we find that the fractions of dark matter and dark energy that are responsible for the interactions (via dark matter decay/annihilation, or other mechanisms) only account for $\sim$2\% and $\sim$1\% which are conveniently derived from $\Delta\lambda_1\Omega_m$ and $\Delta\lambda_2\Omega_d$. Therefore, the values of $\lambda_1\sim 0.07$ and $\lambda_2\sim 0.02$ are small in terms of the interacting fractions.

To investigate these hypothetical models ($\lambda_1=0.07$ and $\lambda_2=0.02$) further, we also performed a few tests as Table \ref{cosmo_imp} shows. Compared to the standard $\Lambda$CDM model, the differences induced by these models are less than 5\% for the age of the universe,  6\% for the amplitude of convergence power spectrum  $C_{\ell}^{\kappa\kappa}$ at $\ell=1000$, and 10\% for the amplitude of equilateral bispectrum $B_{\ell}^{\kappa\kappa\kappa}$ at $\ell=1000$. These numbers in the table clearly verify that those hypothetical models ($\lambda_1=0.07$ and $\lambda_2=0.02$) are very small variations of the standard model in terms of their cosmological implications.

Figs. (\ref{fig.Fisher}) and (\ref{fig.Fisherwa}) show the constraints from the Fisher matrix analysis for two forms of equation of state, respectively. The $1\sigma$ and $2\sigma$ contours for any parameter pairs are plotted in each figure. As seen from Fig. (\ref{fig.Fisher}), there are strong degeneracies between parameter of the equation of state $w_0$ and the parameters of interaction $\lambda_1$, $\lambda_2$. We also find that the two coupling parameters $\lambda_1$ and $\lambda_2$ are highly correlated because they affect the lensing signal in a similar way as $Q=3\lambda_1H\rho_c+3\lambda_2H\rho_d$ indicates. Moreover, the degeneracies among other parameters are also seen from the plot, e.g., $\Omega_{c0}\mbox{--}\sigma_8$. The degeneracy between $w_0$ and $w_a$ seen from Fig. (\ref{fig.Fisherwa}) is caused by the relation $w=w_0+w_a(1-a)$. The evolving parameter $w_a$ is strongly correlated with the coupling parameters $\lambda_1$ and $\lambda_2$, whereas the degeneracy between $w_0$ and $\lambda_1$, $\lambda_2$ are slightly weakened compared to the constant equation of state in Fig. (\ref{fig.Fisher}). To break the degeneracies and tighten the constraints, a joint Fisher matrix analysis with external data sets, such as CMB, SNIa, could be performed.

\section{Conclusions}

In this work we have focused on the models with interactions between dark matter and dark energy, and investigated the constraining power using weak lensing  bispectrum tomography. We have made several assumptions in our analysis: (i) baryons and radiation are neglected since their contributions are small in the late time acceleration; (ii) the dark energy is considered as a homogeneous field and any perturbations to the structure formation are neglected since they are very small compared to the dark matter perturbations; (iii) the halofit method is used to estimate the nonlinear matter power spectrum.

We have extended the study of dark matter decay model \cite{Schafer2008} to more general phenomenological models with interactions between dark sectors. We have shown how different forms of the interaction and equation of state can affect the density perturbations and the lensing weighting functions. Furthermore, we have revealed that the matter density perturbations and lensing weighting functions are two competing factors in the weak lensing power spectrum and bispectrum, and the form of interaction and the equation of state determines which factor is dominant. Compared to the convergence power spectrum, it is clear that the bispectrum can give more stringent constraints on the interactions between dark sectors due to its higher signal-to-noise ratio.

We have used the Fisher matrix formalism to investigate the constraining power of a two-bin bispectrum tomography. The constraints on the cosmological parameters are $\Delta\Omega_c\simeq 0.074$, $\Delta\sigma_8\simeq 0.016$ and $h\simeq 0.14$. For the parameters of equation of state , we have $\Delta w_0\simeq 0.095$ if $w$ is assumed to be a constant, $\Delta w_0\simeq 0.1$ and $\Delta w_a\simeq 0.8$ if $w$ is time-dependent. The errors of the coupling parameters are $\Delta\lambda_1\simeq 0.07$ and $\Delta\lambda_2\simeq 0.02$ which can be further reduced if $w$ is a set to be a constant. The coupling parameters are strongly degenerate with the equation of state, and naturally degenerate with other cosmological parameters. External data sets can be used to break these degeneracies.

Our results show that the weak lensing bispectrum tomography is a sensitive probe to constraining the interaction models. More precision measurements of the cosmic shear in the future will further reduce the parameter space and shed light on the mechanism of the interactions between dark sectors.

\begin{acknowledgments}
CF acknowledges support from NASA grants NASA NNX16AJ69G, NASA NNX16AF39G and Ax Foundation for Cosmology at UC San Diego. This work was partially supported by National Basic Research Program of China (973 Program 2013CB834900)
and National Natural Science Foundation of China.

\end{acknowledgments}

\end{document}